%% file: ds1562.tex
\documentstyle[epsf]{l-aa}  % IRA version
\def\phast{\phantom{$^\ast$}}
\def\ph1{\phantom{1}}
\def\phu{\phantom{.0}}
\begin{document}
\thesaurus{13.18.2 - 11.07.1 - 11.17.3}
\title{Multi-Frequency Study of the B3--VLA Sample.\\
I. 10.6-GHz Data}
\author{L. Gregorini\inst{1,2} \and M. Vigotti\inst{1} 
 \and K.-H. Mack\inst{1,3} \and J. Z\"onnchen\inst{3} \and  U. Klein\inst{3}}
\institute{Istituto di Radioastronomia del CNR, Via Gobetti 101, I-40129, 
Bologna, Italy
\and Dipartimento di Fisica, Universit\`a di Bologna, Via Irnerio 46, I-40126, 
Bologna, Italy
\and Radioastronomisches Institut der Universit\"at Bonn, Auf dem H\"ugel 71,
D-53121 Bonn, Germany }
\offprints{L. Gregorini, Istituto di Radioastronomia del CNR, Via Gobetti 101, I-40129, 
Bologna, Italy}
\date{August 1998}
\maketitle
%
%\markboth{Gregorini et al.}{ }
\begin{abstract}
We present radio continuum data for 1050 B3 radio sources at 10.6~GHz. These 
sources constitute the B3-VLA sample which is complete down to 100~mJy at 408~MHz.
The aim is the construction of a homogeneous spectral database for a
large sample of radio sources, 10 times fainter than the K\"uhr et al.
(1981) sample, in the range 151~MHz to 10.6~GHz. 
Extended and complex radio sources (53) were mapped; the remaining ones were 
observed with cross-scans. We detected 99\% of the radio sources with a flux
density error of about 1 mJy for the fainter ones. The analysis of the quality of the 
10.6~GHz data is presented.
\keywords{Radio continuum: general -- Galaxies: general -- Quasars: general}
\end{abstract}
\section{Introduction} 

The study of large samples of radio sources over a frequency range as wide
as possible is still an important branch of modern astrophysics because of
its cosmological relevance. While early radio surveys naturally discovered
strong sources (many of which are also intrinsically luminous), the need
for surveys of intermediate or low flux densities became apparent. One
important radio survey aiming at lower source flux densities was the 
so-called B3 survey (Ficarra et al. 1985), carried out at 408~MHz and 
complete down to 100~mJy at that frequency and naturally delivering a huge 
number of radio sources, in this case 13354. A subsample of 1050 sources 
was later observed with the VLA (hereafter B3-VLA sample; Vigotti et al. 
1989), the selection criterion ensuring roughly equal numbers of sources in 
5 flux density intervals. Measurements of 429 sources at 1.4~GHz and of 770 
sources at 4.75~GHz were conducted by Kulkarni et al. (1990). Based on these 
surveys the spectral properties and their possible spectral evolution has 
been investigated and discussed (e.g. Kulkar\-ni \& Mantovani 1985; Kapahi 
\& Kulkar\-ni 1986).
\begin{figure}[t]
%\rotate[r]{
\epsfxsize=9cm
\epsfbox  [45 190 540 630] {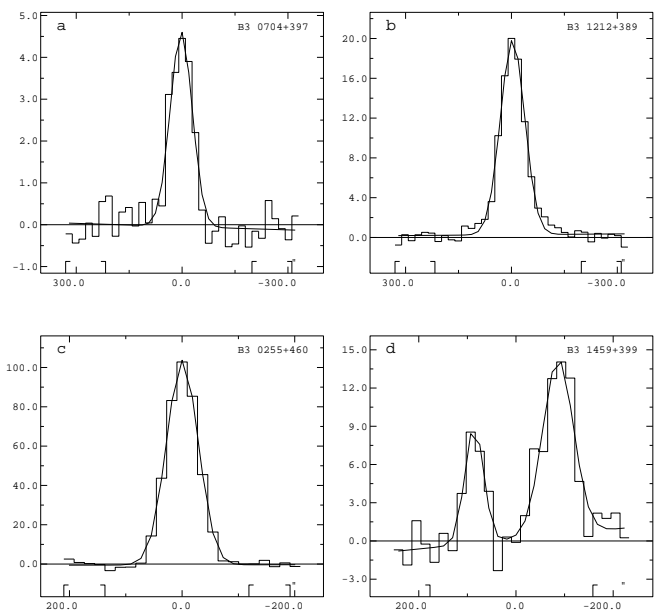}
%}
%\picplace{1 cm}
\caption[]{Examples of cross-scans at $\lambda$2.8~cm: 
a) a 4 mJy source, b) a 20 mJy source, c) a 100 mJy source, d) a double source}
\end{figure}

Our aim is the construction of a homogeneous spectral database for a large 
radio sources sample 10 times fainter than that of K\"uhr et al. (1981).
The 6C survey (Hales et al. 1988) contributes the 151-MHz flux density, the WENSS 
(Rengelink et al. 1997) the 327-MHz flux density. The 408-MHz flux density
is taken from the B3 survey 
(Ficarra et al. 1985) while the NVSS (Condon et al. 1998) yields 
the 1400-MHz flux density. The GB6 survey (Gregory et al. 1996) combined with 
the work of Kulkarni et al. (1990) provides the 4.75~GHz flux densities for 
80\% of the B3-VLA sample.

\include{ds1562t1.tab}
We have therefore embarked on a project to survey the entire sample at 10.6~GHz 
(i.e. $\lambda$2.8~cm) using the Effelsberg 100-m telescope. 
Since all measurements were carried out using IF polarimeters we have
information about the linear polarization of a large number of sources, which
will be reported at a later stage.
\begin{figure}[t]
%\rotate[r]{
\epsfxsize=8cm
\epsfbox  [20 190 520 700] {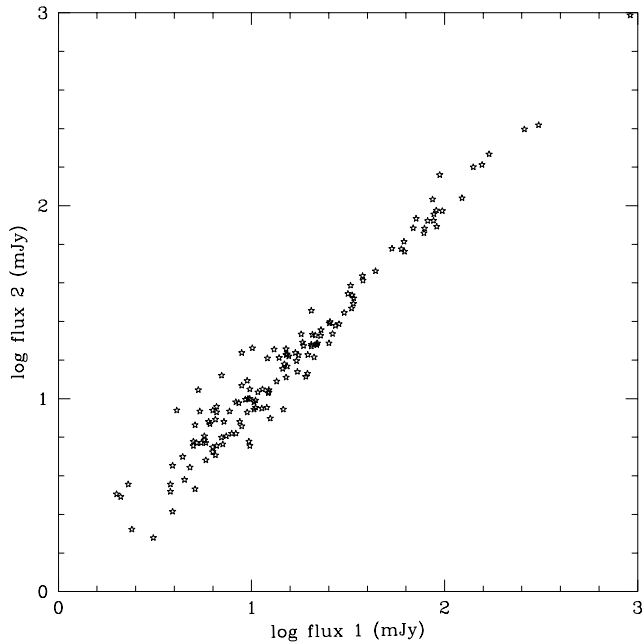}
%}
%\picplace{1 cm}
\caption[]{Plot of flux densities measured at two different epochs.}
\end{figure}

In Sect.~2 we describe the observing techniques and data reduction. Section~3 
presents the results. 

\section{Observations and data reduction}

The observations reported here were carried out between May 1994 and 
January 1996. The 4-feed receiver system installed in the secondary focus
of the 100-m telescope was used in a multi-beam mode. Each horn feeds
a 2-channel receiver with an IF polarimeter providing full Stokes information
simultaneously. The system temperature was $\sim$80~K on the sky (zenith), the 
effective bandwidth was 300~MHz. In the beginning of 1995 the receiver's band 
centre had to be moved from 10.55~GHz to 10.45~GHz in order to avoid the new 
ASTRA~1D satellite. This is only a change of 1\% in frequency, which 
will not have any noticeable influence on the observed source properties (for 
a source with spectral index $\alpha = -1$ this implies a 1\% change in flux 
density; S$_{\nu} \sim \nu^{\alpha}$). The nominal half-power beam width is 
69$^{\prime\prime}$. The total number of sources observed was 1050.

\begin{figure}[t]
\epsfxsize=9cm
\epsfbox  [65 190 560 688] {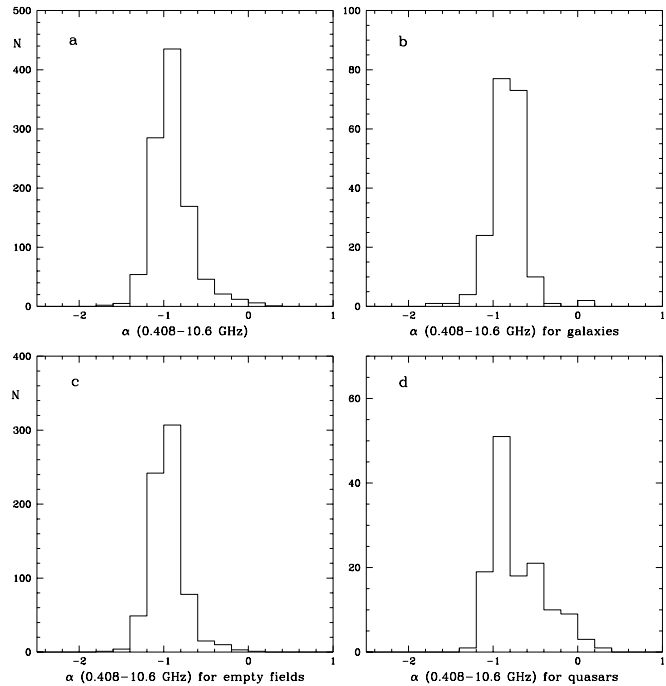}
\caption[]{Histograms of spectral indices between 408~MHz and 10.6~GHz. 
a) for the whole sample, b) for POSS-I galaxies, c) for empty 
fields, and d) for quasars (see text).}
\end{figure}

\subsection{Cross-Scans}

Almost all the sources ($95\%$) were observed with cross-scans, with the main 
beam scanning a distance of 7$^{\prime}$ to ensure adequate baselines. The 
scanning speed was 30$^{\prime}$/min. The 
offset feeds were used to efficiently remove atmospheric noise. For sources 
less extended than 30$^{\prime\prime}$ the cross-scans were oriented in right 
ascension and declination. In the case of more extended sources, 
the cross-scans 
were oriented with one scan direction along the sources' major axes 
(e.g. along double or triple components), with the scan length increased to 
10\arcmin. This orientation was taken from the VLA maps of Vigotti et al. 
(1989). Depending on the expected flux density, the total number of such 
cross-scans was chosen to be between 8 and 64. Individual subscans were 
checked for interference or residual atmospheric fluctuations and
discarded prior to averaging if necessary.
We evaluated the differential signals between the main horn and two of the 
reference horns, which have beam throws of +3$^{\prime}$ and $-5^{\prime}$ 
in azimuth. This allowed a proper judgement of the data quality and enabled 
us to recognize confusing sources that had been accidentally scanned across 
by the reference feeds, thus causing a negative response: the probability of 
picking up unrelated background sources in both reference beams simultaneously 
is very small so that such a negative response only shows up
in one of the 
two recorded differential signals. The data in the final cross-scans were 
sampled at 18$^{\prime\prime}$~intervals.
 With the above scanning speed, 
this implies an integration time of 0.6~second in each subscan. 
Averaging also the two scanning directions we obtain a nominal
rms noise in the final cross-scans between $\sim$2.4~mJy/b.a. (8 scans) 
and $\sim$0.9~mJy/b.a. (64 scans). 
The actually measured rms noise values were generally somewhat higher 
because of residual atmospheric noise. A fit with one Gaussian was applied to 
the final scan yielding the amplitude, the width, and the position of the 
centroid of the Gaussian for Stokes parameters I, Q and U. For the double 
sources with diameters larger than 40$^{\prime\prime}$ we applied a fit with 
two Gaussians only to the scan along the major axis of the radio source. The 
decomposition was successful for 74 sources, whose data have S/N larger than 
10. As mentioned in the introduction, the linear polarization of the sources 
will be presented in a future paper. Figure~1 displays some template plots of 
the cross-scans for sources with different flux densities.

\begin{table}
\caption[]{Spectral indices for different optical identifications}
\begin{tabular}{lccc}
                    &      N     & $\overline{\alpha}$ &  $\Delta \alpha$ \\\\ 
  Galaxies on POSS-I &    181     &    $-0.837$     &   0.021  \\
   Quasars          &    132     &    $-0.716$     &   0.077  \\
   Empty Fields     &    701     &    $-0.969$     &   0.007  \\
\end{tabular}
\end{table}

Standard calibration sources were cross-scanned at regular intervals (about 
every two hours, with two cross-scans each) to check the telescope pointing 
and flux density scale. For the latter purpose the primary calibrators were 
3C\,286 and 3C\,295, with 3C\,48 and 3C\,138 being used as secondaries. 
The pointing 
accuracy was found to be stable to within $\sim3^{\prime\prime}$, sufficiently 
good to ensure reliable flux density measurements. The exact flux density scale 
for each target source was applied by checking two subsequent observations of 
calibration sources. The calibrated flux densities are on the flux density 
scale of Baars et al. (1977).

In order to recover the 
total flux the source extension can be used. Therefore we computed the FWHM
obtained from the Gaussian fit of the data for point-like sources obtaining 
the following results: a mean value of $70^{\prime\prime}\pm4^{\prime\prime}$ 
for sources stronger than 50 mJy, and for fainter sources 
$71^{\prime\prime}\pm10^{\prime\prime}$. The spread of FWHM found permits to 
recover the total flux with an error of up to 15\%. We then decided to determine
the integrated flux using a simulation program. 
The correction factor has been applied to all double and diffuse sources with 
an angular extent between 20$^{\prime\prime}$ and 40$^{\prime\prime}$ and, 
in addition, to more extended sources where the deconvolution could not 
be done because of low S/N. 

Our simulation program was built using two point-like components with a flux
density ratio R$_{20}$ obtained from the VLA 1.4-GHz maps convolved with the
Effelsberg beam (HPBW = 69$^{\prime\prime}$). We had to use R$_{20}$ instead 
of the unknown flux density ratio of source components at 10.6 GHz (R$_{2.8}$);
however, Fig.~4b (see below) will show that R$_{2.8}$ changes by up to a factor
of 2. Our simulation shows that this introduces an additional average error of less than 4$\%$.

Another simulation, which takes into account the extended brightness 
distribution, was used to compute the correction factor for the 
diffuse sources. The factor for double and diffuse sources is in the range 
between $\sim$10$\%$ and $\sim$35$\%$.
\begin{figure}[h]
\epsfxsize=8cm
\epsfbox  [70 335 500 550] {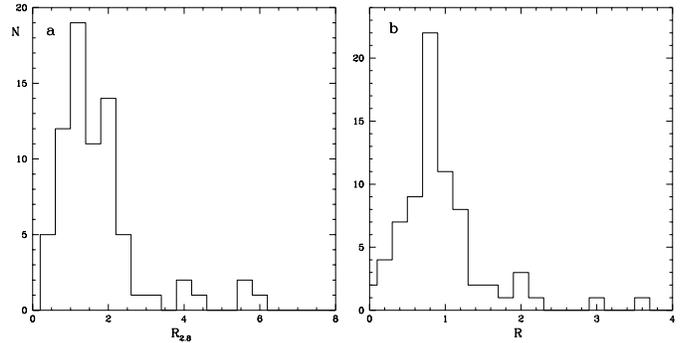}
\caption[]{Histograms of ratios of component flux densities of double 
sources at a) $\lambda$2.8~cm, and b) the ratio of the ratios (see text)} 
\end{figure}
\subsection{Maps}

A total of 53 sources were mapped: they have complex structures larger 
than 70$^{\prime\prime}$. The map sizes were adjusted such as to ensure 
sufficient baseline areas and accounting at the same time for the maximum 
beam throw of the four-feed system, which is 17$^{\prime}$. The standard 
mapping technique with this receiver system was described by Gregorini et 
al. (1992). All four horns were employed by observing in the multi-beam 
mode and applying the restoration algorithm of Emerson et al. (1979). 
Depending on the strength of the sources, between 2 and 14 coverages were 
obtained by scanning the telescope in azimuth and separating subsequent 
scans in elevation by 20$^{\prime\prime}$. After restoration to the
equivalent  
single-beam map the individual coverages were averaged (in I, Q, and U) 
and then interpolated onto a grid in right ascension and declination. Also 
in the course of the mapping campaigns standard calibrators were
observed and processed in the same way as the target sources. All maps 
were numerically integrated to yield total flux densities which were also 
brought to the flux density scale of Baars et al. (1977) by comparison with 
the mapped flux calibrators. Some of the maps show considerable detail.
These will be displayed and briefly described in Sect.~3. 
\section{Results}

\subsection{Flux densities at $\lambda$2.8~cm}
The flux densities derived as described in Sect.~2 are compiled in Tab.~1.
Column~1 gives the B3 source names, Cols.~2 and 3 the radio centroid (equinox
B1950.0) from Vigotti et al. (1989) (computed as the geometric mean
\include{ds1562t3.tab}
\include{ds1562t4.tab}
of the
source components). Columns~4 and 5 contain the measured peak (${\rm S}_{\rm
peak}$) and integrated (${\rm S}_{\rm int})$ 
flux densities. The peak flux density was not given for sources that were
mapped. When a source was not detected we give an 
upper limit (marked with a
`$<$' in Col. 5), which corresponds to a 3-$\sigma$ noise computed for the final
cross-scan. Column~6 is the updated optical identification. 
The symbols are 
g: radio galaxy identified on the POSS-I, most of which are at 
${\rm z} \le 0.5$; G: far 
radio galaxy with measured redshift ($0.5 \le {\rm z} \le 3.5$); Q:
spectroscopically confirmed quasar; b: blue objects (i.e. non-confirmed
quasars); BL: BL Lac; F: featureless spectrum; a blank means `empty field', i.e.
it lacks any optical counterpart down to the POSS-I limit (more than 90$\%$ are
distant radio galaxies, the remaining ones being quasars with magnitudes fainter
than the POSS-I). So far the B3-VLA sample is composed of $27\%$ galaxies,
$12\%$ quasars, and $61\%$ empty fields.
\begin{figure}
\epsfxsize=8cm
\epsfbox  [30 180 534 680] {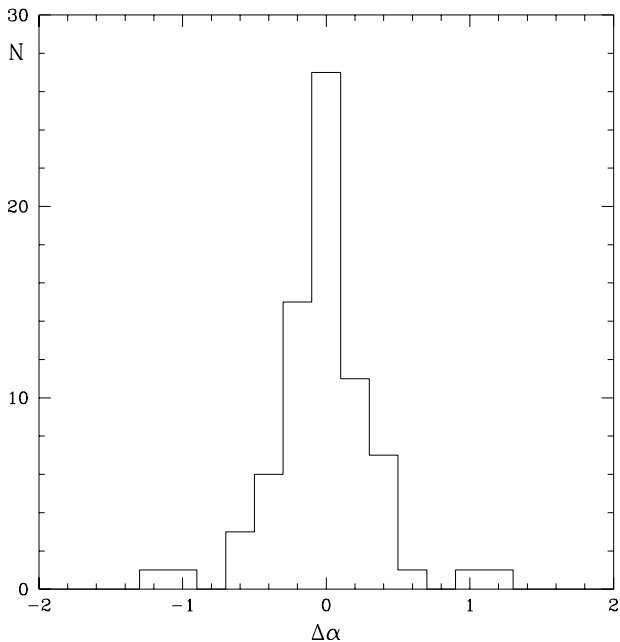}
\caption[]{Histogram of the difference $\Delta\alpha$ between the spectral 
indices of the two components of double sources}
\end{figure}
In order to have an independent check on the accuracy of our measurements 
and to evaluate the uncertainties of the flux densities, 143 sources were
re-observed. These sources were chosen in different flux bins. Figure~2
shows the plot of the fluxes measured in the two independent observations. 
Intrinsic source variability increases the scatter of the plot. The analysis 
of the distribution of the differences between two or more independent flux 
measurements of the same source allowed for a determination of the random 
errors affecting the measurements. These errors are the quadratic sum of 
three terms: the first, proportional to the source intensity, is introduced 
by gain instabilities of the receiver; the other two, independent of the 
source flux density, are due to noise and confusion. The first term was 
computed using sources with flux densities greater than 500~mJy as well 
as calibrators, and was found to be $\sim 2\%$. From the faint sources we 
evaluated the rms noise; we subdivided the sources into different classes 
according to number of scans (i.e. observing time) and found that this term 
is in the range 0.6--0.8 mJy. To be conservative we chose 0.8 mJy. For the 
confusion term we refer to Reich (1993) who reports a 
value of 0.08 mJy. We expect that the error affecting the flux density
measurements is:

$$\sigma_{\rm S} = \sqrt{(0.02 \cdot \rm S_{\rm tot})^2 + 0.08^2 + 0.8^2}$$

\noindent
where $\rm S_{\rm tot}$ is in mJy. To compute the spectral index between 
408~MHz and 10.6~GHz (presented in Fig.~3a) the low-frequency flux densities 
were 
increased by 5\% to adjust them to the scale of Baars et al. (1977). The 
spectral indices of B30226+394, B30241+393B, B30920+408, 
B31016+388B, B31428+385, B31447+402, B32333+397, and B32348+387 were 
not included; the VLA maps 
show that the components of these triple sources are probably not physically 
connected.

\begin{figure*}
\epsfxsize=14.5cm
\epsfbox  [-10 80 514 840] {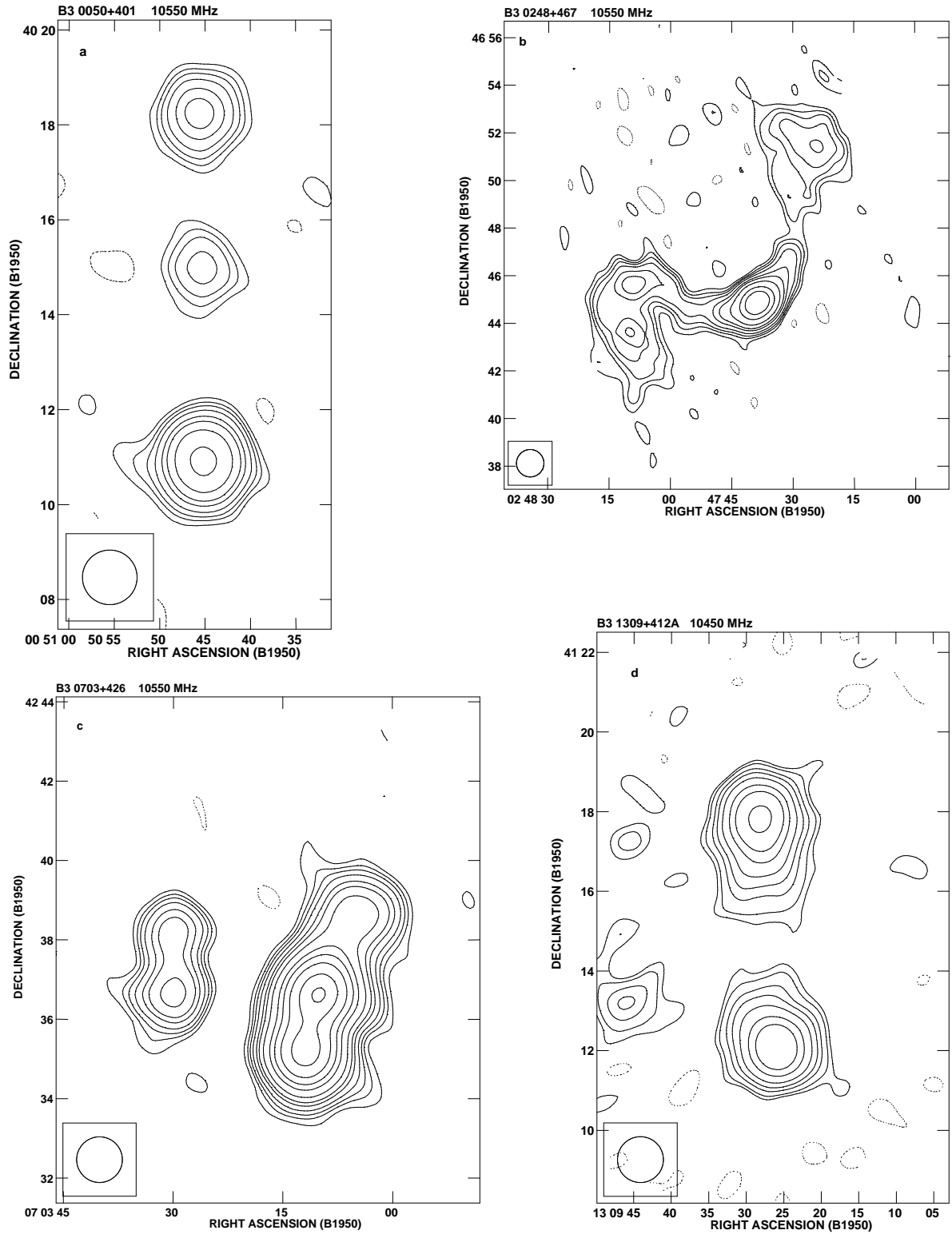}
\caption[]{Maps of four B3 sources with complex structure at $\lambda$2.8~cm:
a) B3\,0050+401 -- contour levels: -3, 3, 5, 7, 10, 15, 20, 30, 50~mJy ; b)
B3\,0248+467 -- contour levels: -3, 3, 5, 7, 10, 15, 20, 30, 50, 70, 100, 150,
200~mJy; c) B3\,0703+426 -- contour levels: -3, 3, 5, 7, 10, 15, 20, 30, 50, 70,
100, 150, 200~mJy; d) B3\,1309+412A -- contour levels: -2, 2, 3, 5, 7, 10, 15,
20, 30~mJy}
\end{figure*}

The spectral indices have a broad distribution with a median value of 
$\alpha_{med}=-0.933$ and an average value of $\overline{\alpha}=-0.906$. 
Figures~3b, 3c, and 3d show the histograms of the 
spectral indices for three different classes of optical counterparts, i.e.  
galaxies bright enough to appear on the POSS-I (most of which are at 
${\rm z} < 0.5$), empty fields on the POSS-I (most of which are galaxies 
${\rm z} > 0.5$), and quasars. In Tab.~2 we have compiled the mean spetral 
indices ($\overline{\alpha}$) and their uncertainties ($\Delta \alpha$)
for different optical identifications.

The distributions of the first two classes 
are similar, 
but the different average values (see Tab.~2) indicate that high-redshift 
radio galaxies have steeper spectra. The distribution for quasars shows the 
presence of two populations: steep-spectrum and flat-spectrum quasars. 

For the decomposed double sources (see Sect.~2) Tab.~3 presents in Col.~1 
the B3 source names, in Cols.~2 and 3 the positions (equinox B1950.0) of the 
components from Vigotti et al. (1989); Cols.~4 and 5 give the flux densities at 
$\lambda$2.8~cm and $\lambda$20~cm (Vigotti et al. 1989), respectively. 
For each source we have 
computed the component ratio R$_{2.8}$ of the flux density at $\lambda$2.8~cm 
(see Fig.~4a), and the ratio R$_{20}$ of the flux density at $\lambda$20~cm, 
then the 
ratio R=R$_{2.8}$/R$_{20}$ (see Fig.~4b). These ratios were computed 
considering as the first component the brightest one at 20~cm. Figure~4a shows 
that the double sources at $\lambda$2.8~cm are symmetric with respect to the 
flux densities; in fact only 18\% have a ratio greater than 2 (very high 
values could 
also be an indication that the two components are not physically connected). 
Figure~4b shows that only 10\% of double sources change their flux density 
ratio between 1.4 and 10.6~GHz by more than a factor of 2.

The spectral index computed between $\lambda\lambda$20~cm and 2.8~cm for the 
single components has a median value of $-0.890$. Figure~5 presents the 
difference
$\Delta\alpha$ between the spectral indices of the two components of the 74 
double sources. The distribution of $\Delta\alpha$ is asymmetric, in the 
sense that source components that are brighter at $\lambda$20~cm exhibit 
slightly steeper spectra on average. 

The parameters of the mapped sources are presented in Tab.~4. Column~1 gives 
the B3 source names, Col.~2 a letter which marks the component. All the 
parameters presented in Cols.~3 through 8 are determined with a 
two--dimensional Gaussian fit to the $\lambda$2.8cm data. Columns~3 and 4 
contain the component positions (equinox B1950.0); Cols.~5 and 6 give the 
integrated flux of the whole source and of the components, respectively; 
Cols.~7 and 8 have the FWHM and position angle. Column~9 gives the noise of 
the map. For complex sources, not well approximated by a Gaussian, the total 
flux densities are marked with an asterisk. Note that the rms noise in the 
maps is 
comparatively low with respect to the cross-scans. This is due to the fact 
that in mapping all four horns have been used, while in the cross-scans only 
the main beam was measuring the target, and the offset feeds were used to 
reject atmospheric noise. It is clear that the reference feeds add 
uncorrelated noise to the differential signal in the cross-scans, whereas 
in case of mapping they contribute to the source flux density measurement. 
In Fig.~6 we show five maps of those sources that exhibit the most complex 
structure at this resolution. 

\subsection{Notes on individual mapped sources}
\begin{itemize}

\item {\it B3\,0050+401}: This is a triple source consisting of a core and 
two lobes. The core has an inverted spectrum ($\alpha = 0.64$) between 
1.465 and 10.6~GHz. Note that there is an error in Tab.~II of Vigotti
et al. (1989), which gives 111~mJy for the core at 1.465~GHz, while this
should be 4~mJy. The source has a total spectral index $\alpha = -0.719$
between 0.408 and 10.6~GHz (Fig. 6a).

\item {\it B3\,0157+393}: This one is reminiscent of wide-angle tailed sources. 
The brightness asymmetry visible at 1.465~GHz (Vigotti et al. 1989) is also 
visible at 10.6~GHz. The spectrum is relatively flat ($\alpha= -0.59$) 
between 408~MHz and 10.6~GHz, indicating that the core may contribute 
proportionally more at the high frequency.

\item {\it B3\,0157+405}: This turns out to be a giant radio galaxy 
(Schoenmakers, priv. comm.). The spectrum is steep ($\alpha = -1.18$) between 
0.408 and 10.6~GHz, indicating significant particle ageing. This is in line 
with the rather relaxed radio continuum morphology seen at boh frequencies, 
and with the lack of any core and jets.

\item {\it B3\,0220+427}: This is 3C\,66: between 408~MHz and 10.6~GHz
the spectrum of the compact source B3\,0219+428A (3C66A) is relatively flat 
($\alpha=-0.432$), the radio source B3\,0220+427A/D is a FRI with 
$\alpha=-0.646$. 

\item {\it B3\,0248+467}: An S-shaped radio galaxy identified with IC260.
At 10.6 GHz the core becomes very prominent. The total spectral index between 
0.408 and 10.6~GHz is $\alpha=-0.57$ (Fig. 6b).

\item {\it B3\,0703+426}: The map contains two sources: B3\,0703+426A and 
B3\,0703+426B. \\
The first, identified with a cluster galaxy, has $\alpha=-0.60$ (between 
0.408 and 10.6~GHz); the second, identified with a compact object, has 
$\alpha=-0.82$ (Fig. 6c). 

\item {\it B3\,1309+412A}: A FRII radio galaxy; the spectral index was not computed 
because the 408~MHz flux density also includes B3\,1309+413 (Fig. 6d).

\item {\it B3\,1450+391}: A very diffuse radio source probably identifiable with 
a cluster galaxy. The spectrum is steep ($\alpha=-0.97$); however, the 
flux densities at 408~MHz and at 10.6~GHz can be underestimated.

\end{itemize}

\begin{acknowledgements}
We are grateful to H. Rottmann and G. Zech for their invaluable help during 
the many observing runs. 
We wish to thank our referee, Dr. J.J. Condon, for his useful comments.
Part of this work was supported by the Deutsche Forschungsgemeinschaft, grant 
KL533/4-2, and by the European Commission, TMR Programme, Research Network 
Contract ERBFMRXCT97-0034 ``CERES''.
We thank Dr. H. Andernach for a careful check of the tables. 
\end{acknowledgements}

\end{document}

%% file: ds1562t1.tab.tex
\begin{table*}
\caption[]{B3\,VLA sources at 10.6 GHz}
\begin{flushleft}
\begin{tiny}
\begin{tabular}{|lccrrc|lccrrc|}
\hline
  B3 name & RA (B1950) & DEC (B1950) & S$_{\rm peak}$ & S$_{\rm int}$ & ID &  B3 name & RA (B1950) & DEC (B1950) & S$_{\rm peak}$ & S$_{\rm int}$ & ID\\
 & [$^{\rm h}$ ~ $^{\rm m}$ ~ $^{\rm s}$] & [\degr$\;\;\;$\arcmin$\;\;\;$\arcsec] & [mJy] & [mJy] & & & [$^{\rm h}$ ~ $^{\rm m}$ ~ $^{\rm s}$] & [\degr$\;\;\;$\arcmin$\;\;\;$\arcsec] & [mJy] & [mJy] &\\\hline
   0000+394  & 00 00 08.3 & 39 29 52 &         &    17.5 &    & 0109+415  & 01 09 16.5 & 41 31 05 &    97.4 &   106.9 &   g\\       
   0000+399  & 00 00 57.7 & 39 56 44 &    12.6 &    12.6 &    & 0109+416B & 01 09 23.5 & 41 39 18 &    59.9 &    59.9 &   g\\       
   0001+395  & 00 01 41.4 & 39 33 10 &     1.5 &     1.5 &    & 0110+386  & 01 10 07.2 & 38 37 23 &    20.3 &    21.8 &    \\       
   0001+398  & 00 01 45.0 & 39 49 27 &     9.1 &    10.9 &   g& 0110+395  & 01 10 46.6 & 39 34 40 &    39.1 &    39.1 &    \\       
   0003+380  & 00 03 22.3 & 38 03 33 &   655.2 &   655.2 &   g& 0110+398  & 01 10 37.6 & 39 54 35 &     5.7 &     5.7 &    \\       
   0003+387  & 00 03 45.8 & 38 43 45 &    29.6 &    29.6 &   G& 0110+401  & 01 10 26.5 & 40 10 19 &   143.6 &   143.6 &   Q\\       
   0004+380A & 00 04 01.4 & 38 02 25 &    27.0 &    27.0 &    & 0112+400  & 01 12 23.5 & 40 04 20 &    17.9 &    17.9 &    \\       
   0005+383B & 00 05 47.4 & 38 20 30 &         &    92.5 &   g& 0112+432  & 01 12 03.5 & 43 16 25 &    41.9 &    41.9 &    \\       
   0006+397  & 00 06 28.5 & 39 45 05 &   253.4 &   253.4 &   Q& 0113+400  & 01 13 32.7 & 40 01 36 &    88.0 &    93.2 &    \\       
   0008+392  & 00 08 01.4 & 39 17 34 &    13.4 &    15.3 &    & 0114+399  & 01 14 35.7 & 39 57 03 &    14.4 &    15.4 &    \\       
   0010+392  & 00 10 07.1 & 39 16 15 &    18.8 &    18.8 &    & 0115+394  & 01 15 03.5 & 39 28 46 &     7.4 &     7.4 &    \\       
   0010+395  & 00 10 35.9 & 39 31 26 &     2.8 &     3.1 &    & 0115+453A & 01 15 03.5 & 45 20 35 &   144.4 &   144.4 &   G\\       
   0010+402  & 00 10 16.5 & 40 16 07 &    49.6 &    49.6 &   g& 0115+469  & 01 15 25.3 & 46 56 35 &         &    64.3 &    \\       
   0010+405  & 00 10 54.2 & 40 34 56 &   818.5 &   818.5 &   g& 0116+397  & 01 16 25.2 & 39 43 06 &     7.0 &     7.0 &    \\       
   0013+387  & 00 13 22.8 & 38 43 47 &    41.5 &    41.5 &   Q& 0116+438  & 01 16 27.5 & 43 51 54 &    60.6 &    70.7 &   g\\       
   0013+393  & 00 13 11.2 & 39 20 27 &    16.1 &    16.1 &    & 0119+395  & 01 19 34.6 & 39 32 46 &     7.9 &     7.9 &    \\       
   0014+395  & 00 14 16.4 & 39 31 23 &     9.2 &     9.2 &    & 0119+397  & 01 19 40.9 & 39 45 26 &    30.5 &    30.5 &    \\       
   0015+399  & 00 15 03.7 & 39 59 49 &    32.0 &    32.0 &    & 0120+380  & 01 20 00.0 & 38 02 41 &     9.6 &    10.3 &    \\       
   0017+395  & 00 17 10.0 & 39 35 15 &     8.6 &     8.6 &    & 0120+405  & 01 20 32.2 & 40 31 20 &    65.4 &    65.4 &   G\\       
   0017+432  & 00 17 29.4 & 43 13 31 &    26.6 &    26.6 &    & 0121+389  & 01 21 15.9 & 38 57 48 &    19.8 &    19.8 &    \\       
   0018+393  & 00 18 44.1 & 39 21 55 &    12.1 &    18.0 &    & 0122+395  & 01 22 34.7 & 39 30 17 &     8.9 &     9.3 &    \\       
   0019+391  & 00 19 21.0 & 39 09 08 &     5.7 &     5.7 &    & 0123+385  & 01 23 17.7 & 38 35 12 &    16.4 &    16.4 &   G\\       
   0019+431  & 00 19 08.6 & 43 11 47 &    66.9 &    74.8 &   Q& 0123+396  & 01 23 31.9 & 39 38 39 &    39.0 &    40.4 &   g\\       
   0020+437  & 00 20 51.3 & 43 46 26 &    44.2 &    44.2 &   F& 0123+402  & 01 23 04.9 & 40 13 01 &    25.0 &    25.0 &    \\       
   0021+383  & 00 21 34.1 & 38 18 33 &    28.7 &    28.7 &    & 0124+387  & 01 24 50.8 & 38 44 39 &    15.6 &    15.6 &    \\       
   0021+395  & 00 21 38.7 & 39 32 29 &     9.6 &     9.6 &    & 0126+392A & 01 26 33.8 & 39 14 27 &     6.8 &     6.8 &    \\       
   0022+390  & 00 22 46.7 & 39 02 59 &   493.1 &   493.1 &   Q& 0127+395  & 01 27 47.6 & 39 34 48 &     6.1 &     6.1 &    \\       
   0022+394  & 00 22 24.4 & 39 29 18 &    12.8 &    12.8 &    & 0127+399  & 01 27 54.3 & 39 54 41 &     8.7 &     8.7 &    \\       
   0022+399  & 00 22 03.4 & 39 59 35 &    21.8 &    21.8 &   Q& 0128+394  & 01 28 34.7 & 39 27 32 &    24.9 &    24.9 &    \\       
   0022+424  & 00 22 54.9 & 42 27 15 &    59.8 &    59.8 &   g& 0130+381  & 01 30 48.6 & 38 07 38 &   105.3 &   105.3 &   g\\       
   0023+382  & 00 23 30.0 & 38 14 55 &    48.1 &    48.1 &   g& 0130+384  & 01 30 05.3 & 38 25 39 &    27.0 &    32.8 &   Q\\       
   0025+394  & 00 25 53.5 & 39 28 06 &     5.1 &     5.5 &   g& 0130+398  & 01 30 17.3 & 39 47 59 &    25.2 &    25.2 &    \\       
   0026+397  & 00 26 50.7 & 39 46 40 &     2.8 &     4.5 &   g& 0131+390  & 01 31 49.8 & 39 04 33 &         &    27.0 &   g\\       
   0027+380  & 00 27 42.1 & 38 00 48 &    22.5 &    22.5 &   Q& 0132+376A & 01 32 35.1 & 37 38 54 &         &   196.7 &   g\\       
   0027+395  & 00 27 03.4 & 39 32 04 &    38.0 &    38.0 &   Q& 0132+392  & 01 32 38.9 & 39 12 54 &    20.9 &    20.9 &    \\       
   0028+390  & 00 28 56.9 & 39 02 31 &    27.0 &    27.0 &    & 0133+381  & 01 33 43.4 & 38 10 45 &    20.5 &    21.4 &    \\       
   0028+394  & 00 28 30.4 & 39 25 23 &    20.9 &    22.9 &    & 0134+386  & 01 34 54.0 & 38 36 13 &    63.0 &    77.6 &   g\\       
   0028+409  & 00 28 07.6 & 40 54 20 &    17.4 &    17.4 &   G& 0134+389  & 01 34 20.1 & 38 56 20 &    17.4 &    17.4 &    \\       
   0028+450  & 00 28 08.4 & 45 05 14 &    32.6 &    38.0 &   g& 0136+396  & 01 36 34.0 & 39 41 56 &   115.0 &   115.0 &   g\\       
   0029+394  & 00 29 54.0 & 39 25 42 &    43.6 &    43.6 &   G& 0137+385  & 01 37 50.5 & 38 33 39 &    15.3 &    19.7 &    \\       
   0029+398  & 00 29 34.9 & 39 50 38 &    57.4 &    57.4 &    & 0137+401  & 01 37 36.3 & 40 09 04 &    49.7 &    49.7 &   Q\\       
   0030+390  & 00 30 13.6 & 39 03 43 &     5.6 &     5.6 &    & 0138+394  & 01 38 31.5 & 39 25 30 &    14.4 &    14.4 &    \\       
   0031+391  & 00 31 33.1 & 39 07 42 &   127.1 &   139.4 &   G& 0139+389A & 01 39 13.0 & 38 57 33 &    42.1 &    42.1 &   Q\\       
   0031+393  & 00 31 30.0 & 39 19 42 &    19.5 &    19.5 &    & 0140+387  & 01 40 36.0 & 38 47 06 &    17.5 &    17.5 &    \\       
   0032+390  & 00 32 29.2 & 39 02 35 &     8.8 &     9.3 &    & 0141+398  & 01 41 33.8 & 39 48 11 &    11.0 &    11.0 &    \\       
   0032+394  & 00 32 43.8 & 39 24 15 &    19.6 &    19.6 &    & 0143+393  & 01 43 02.4 & 39 18 15 &     6.6 &     6.6 &    \\       
   0032+423  & 00 32 23.2 & 42 21 47 &    60.0 &    60.0 &   Q& 0143+446B & 01 43 45.4 & 44 40 25 &    36.5 &    36.5 &   Q\\       
   0033+397  & 00 33 19.3 & 39 45 17 &     5.9 &     5.9 &    & 0144+391  & 01 44 24.5 & 39 09 52 &    13.2 &    13.2 &    \\       
   0033+425  & 00 33 55.3 & 42 35 54 &    19.4 &    21.2 &    & 0144+399  & 01 44 17.9 & 39 58 52 &     9.4 &     9.4 &    \\       
   0034+387  & 00 34 37.1 & 38 42 46 &    22.4 &    22.4 &    & 0144+430  & 01 44 55.8 & 43 04 47 &    31.8 &    32.5 &   Q\\       
   0034+393  & 00 34 54.2 & 39 21 42 &    56.7 &    56.7 &   Q& 0144+432  & 01 44 53.6 & 43 17 19 &    50.6 &    50.6 &   Q\\       
   0034+444  & 00 34 09.6 & 44 26 51 &    68.7 &    68.7 &   G& 0146+394  & 01 46 47.1 & 39 24 20 &    15.1 &    15.1 &    \\       
   0035+385A & 00 35 04.7 & 38 31 25 &         &   126.9 &   F& 0147+397  & 01 47 16.1 & 39 44 46 &    20.8 &    22.3 &   g\\       
   0036+398  & 00 36 49.8 & 39 52 07 &     9.7 &     9.7 &    & 0147+398  & 01 47 36.9 & 39 52 23 &     9.8 &    11.1 &   g\\       
   0037+394  & 00 37 36.3 & 39 28 23 &    12.2 &    12.2 &    & 0147+400  & 01 47 20.1 & 40 02 38 &   162.7 &   162.7 &    \\       
   0037+396  & 00 37 34.0 & 39 38 37 &         &   <5\phu&    & 0149+398  & 01 49 08.7 & 39 49 46 &    72.7 &    79.4 &   g\\       
   0038+399  & 00 38 41.2 & 39 56 04 &    11.6 &    14.6 &   g& 0150+406  & 01 50 48.2 & 40 41 09 &    24.3 &    27.2 &    \\       
   0039+373  & 00 39 24.3 & 37 23 10 &    89.3 &    89.3 &   G& 0152+382  & 01 52 18.8 & 38 16 38 &    47.3 &    59.2 &    \\       
   0039+391  & 00 39 11.6 & 39 08 54 &    23.7 &    23.7 &   G& 0152+435  & 01 52 26.3 & 43 31 20 &   251.6 &   251.6 &   G\\       
   0039+398  & 00 39 33.7 & 39 53 23 &    85.7 &    85.7 &    & 0153+417  & 01 53 18.6 & 41 47 49 &   100.5 &   100.5 &   g\\       
   0039+412  & 00 39 34.5 & 41 13 01 &    61.1 &    61.1 &    & 0157+393A & 01 57 08.6 & 39 19 24 &    11.6 &    13.1 &   G\\       
   0040+470  & 00 40 37.3 & 47 00 12 &    41.3 &    41.3 &   G& 0157+393B & 01 57 50.0 & 39 20 19 &         &   163.5 &   g\\       
   0041+382A & 00 41 05.9 & 38 13 45 &    17.8 &    17.8 &    & 0157+405A & 01 57 27.8 & 40 34 23 &         &    94.6 &   g\\       
   0041+393  & 00 41 14.5 & 39 21 31 &    25.3 &    25.3 &    & 0157+442  & 01 57 32.7 & 44 12 46 &   147.4 &   147.4 &   Q\\       
   0041+405  & 00 41 10.0 & 40 30 10 &    36.7 &    36.7 &    & 0158+391  & 01 58 50.0 & 39 09 57 &    16.2 &    17.5 &    \\       
   0041+425  & 00 41 53.9 & 42 31 37 &    55.5 &    55.5 &    & 0158+394  & 01 59 00.2 & 39 28 54 &   144.3 &   144.3 &   Q\\       
   0042+381A & 00 42 24.7 & 38 06 53 &    70.8 &    70.8 &    & 0159+390  & 01 59 51.6 & 39 02 15 &    20.1 &    20.1 &   g\\       
   0042+381B & 00 42 33.7 & 38 08 52 &    40.0 &    40.7 &    & 0159+397  & 01 59 18.2 & 39 45 12 &     9.9 &    10.1 &    \\       
   0042+386  & 00 42 01.4 & 38 40 36 &    17.0 &    17.0 &    & 0200+393  & 02 00 48.6 & 39 18 19 &    49.7 &    49.7 &    \\       
   0043+398  & 00 43 11.1 & 39 50 19 &     3.9 &     4.9 &   g& 0201+390  & 02 01 43.0 & 39 05 04 &    15.1 &    15.1 &    \\       
   0045+393  & 00 45 17.1 & 39 21 05 &    76.6 &    76.6 &    & 0201+396  & 02 01 36.0 & 39 40 58 &     4.3 &     4.3 &    \\       
   0045+395  & 00 45 10.4 & 39 32 37 &    49.2 &    49.2 &   BL& 0201+402 & 02 01 36.3 & 40 15 21 &    23.2 &    23.2 &    \\       
   0045+396  & 00 46 00.0 & 39 37 20 &    21.4 &    22.4 &   g& 0202+380  & 02 02 52.3 & 38 01 25 &    42.0 &    42.0 &   g\\       
   0045+400  & 00 45 27.6 & 40 05 32 &    60.5 &    67.0 &   g& 0205+395  & 02 05 04.5 & 39 30 47 &    34.4 &    34.4 &    \\       
   0045+404  & 00 45 58.6 & 40 28 43 &    28.3 &    28.3 &    & 0205+398  & 02 05 52.9 & 39 52 50 &     3.4 &     3.4 &    \\       
   0046+439  & 00 46 03.6 & 43 57 22 &    22.6 &    23.7 &   G& 0207+389  & 02 07 06.9 & 38 57 15 &    18.5 &    19.2 &   g\\       
   0049+379  & 00 49 31.5 & 37 59 11 &    56.5 &    56.5 &    & 0207+395  & 02 07 07.3 & 39 35 52 &    19.3 &    19.3 &   Q\\       
   0049+395  & 00 49 27.6 & 39 35 06 &     4.7 &     4.7 &    & 0207+397  & 02 07 52.4 & 39 47 32 &    30.1 &    30.1 &   g\\       
   0050+401  & 00 50 45.3 & 40 14 50 &         &   117.4 &   g& 0207+399  & 02 07 12.6 & 39 59 26 &    13.4 &    13.4 &    \\       
   0051+397  & 00 51 13.6 & 39 42 44 &     8.1 &     8.1 &    & 0209+386  & 02 09 41.0 & 38 36 50 &    22.6 &    23.0 &    \\       
   0051+404  & 00 51 40.5 & 40 25 56 &   111.0 &   111.0 &    & 0209+390  & 02 09 47.2 & 39 03 17 &     5.5 &     5.5 &    \\       
   0052+380  & 00 52 10.1 & 38 05 35 &    30.9 &    32.4 &    & 0209+394  & 02 09 33.8 & 39 28 15 &    42.4 &    42.4 &   Q\\       
   0052+392  & 00 52 53.8 & 39 17 22 &    15.1 &    16.2 &    & 0210+396  & 02 10 30.1 & 39 40 07 &     7.9 &     7.9 &    \\       
   0052+395  & 00 52 53.8 & 39 32 02 &    24.0 &    24.0 &    & 0211+393  & 02 11 02.4 & 39 19 01 &         &    28.2 &   g\\       
   0053+394  & 00 53 02.5 & 39 28 47 &    23.6 &    26.3 &   g& 0213+392  & 02 13 09.0 & 39 17 00 &    14.2 &    14.2 &   G\\       
   0053+439  & 00 53 57.2 & 43 57 11 &    54.0 &    55.1 &    & 0213+398  & 02 13 18.2 & 39 51 39 &     9.7 &     9.7 &   G\\       
   0054+396  & 00 54 42.1 & 39 41 05 &    13.7 &    13.7 &    & 0213+407  & 02 13 46.1 & 40 47 02 &    18.3 &    18.3 &   g\\       
   0056+389  & 00 56 51.2 & 38 54 12 &     7.6 &     7.6 &    & 0213+412  & 02 13 23.8 & 41 17 58 &   116.4 &   116.4 &   G\\       
   0057+395  & 00 57 17.5 & 39 33 26 &    15.0 &    15.0 &    & 0214+393  & 02 14 33.7 & 39 22 28 &         &    85.5 &   g\\       
   0057+397  & 00 57 03.2 & 39 42 37 &         &   <3\phu&    & 0216+388  & 02 16 48.6 & 38 52 00 &     7.3 &     7.3 &    \\       
   0058+403  & 00 58 04.8 & 40 19 52 &    32.6 &    32.6 &    & 0216+393  & 02 16 30.4 & 39 18 51 &    34.5 &    34.5 &    \\       
   0059+397  & 00 59 38.3 & 39 48 14 &     7.5 &     8.3 &    & 0216+403  & 02 16 57.3 & 40 21 58 &    36.1 &    36.1 &    \\       
   0059+461  & 00 59 54.4 & 46 08 31 &    96.6 &   111.8 &    & 0216+423  & 02 16 01.0 & 42 19 13 &   117.2 &   117.2 &    \\       
   0100+388  & 01 00 55.1 & 38 48 35 &    20.2 &    20.2 &    & 0217+395  & 02 17 03.1 & 39 30 56 &     9.8 &     9.8 &    \\       
   0100+397  & 01 00 09.5 & 39 43 56 &     5.9 &     6.0 &    & 0217+417  & 02 17 06.6 & 41 43 51 &    55.0 &    55.0 &   Q\\       
   0103+422  & 01 03 51.5 & 42 13 53 &    76.5 &    76.5 &    & 0218+396  & 02 18 42.7 & 39 41 53 &   156.9 &   156.9 &    \\       
   0105+441  & 01 05 17.8 & 44 09 01 &    78.5 &    78.5 &    & 0218+399A & 02 18 40.7 & 39 55 24 &    20.4 &    20.4 &    \\       
   0106+380  & 01 06 36.5 & 38 00 46 &    40.4 &    40.4 &   Q& 0218+402A & 02 18 55.5 & 40 14 36 &    21.9 &    21.9 &    \\       
   0106+397  & 01 06 35.1 & 39 44 04 &    14.3 &    14.3 &    & 0219+397  & 02 19 33.3 & 39 47 02 &    18.8 &    18.8 &    \\       
   0107+397  & 01 07 00.8 & 39 42 30 &    24.7 &    24.7 &    & 0219+421  & 02 19 24.1 & 42 07 17 &   152.0 &   152.0 &   g\\       
   0107+398  & 01 07 53.9 & 39 51 25 &    13.3 &    13.3 &   Q& 0219+428A & 02 19 30.1 & 42 48 32 &         &  1235.2 &   BL\\       
   0107+399  & 01 07 13.2 & 39 55 16 &     5.8 &     5.8 &    & 0219+443  & 02 19 06.1 & 44 19 18 &    64.1 &    64.1 &   Q\\       
   0108+396  & 01 08 11.4 & 39 36 10 &    10.3 &    10.9 &    & 0220+388  & 02 20 04.4 & 38 49 31 &    12.7 &    12.7 &    \\       
   0108+402  & 01 08 14.1 & 40 17 47 &    21.4 &    21.8 &    & 0220+393A & 02 20 28.2 & 39 22 24 &   165.3 &   169.7 &   g\\       
   0109+390  & 01 09 45.5 & 39 02 10 &     6.9 &     7.4 &    & 0220+397  & 02 20 36.6 & 39 47 17 &   284.6 &   289.5 &   G\\\hline       
\end{tabular}
\end{tiny}
\end{flushleft}
\end{table*}

\setcounter{table}{0}
\begin{table*}
\caption[]{B3\,VLA sources at 10.6 GHz (cont'd)}
\begin{flushleft}
\begin{tiny}
\begin{tabular}{|lccrrc|lccrrc|}
\hline
  B3 name & RA (B1950) & DEC (B1950) & S$_{\rm peak}$ & S$_{\rm int}$ & ID &  B3 name & RA (B1950) & DEC (B1950) & S$_{\rm peak}$ & S$_{\rm int}$ & ID\\
 & [$^{\rm h}$ ~ $^{\rm m}$ ~ $^{\rm s}$] & [\degr$\;\;\;$\arcmin$\;\;\;$\arcsec] & [mJy] & [mJy] & & & [$^{\rm h}$ ~ $^{\rm m}$ ~ $^{\rm s}$] & [\degr$\;\;\;$\arcmin$\;\;\;$\arcsec] & [mJy] & [mJy] &\\\hline
   0220+427A & 02 20 06.8 & 42 46 02 &         &  1806.3 &   g& 0726+431  & 07 26 16.7 & 43 07 35 &    48.3 &    50.4 &   Q\\       
   0221+383  & 02 21 12.2 & 38 18 34 &    28.8 &    28.8 &    & 0727+401  & 07 27 10.0 & 40 07 42 &    49.3 &    49.3 &    \\       
   0221+393  & 02 21 37.3 & 39 19 14 &         &     7.6 &    & 0728+389  & 07 28 34.4 & 38 57 51 &    22.0 &    22.0 &    \\       
   0221+396  & 02 21 44.4 & 39 41 44 &    27.2 &    32.6 &    & 0728+395  & 07 28 30.8 & 39 29 59 &     9.3 &     9.6 &    \\       
   0222+403  & 02 22 37.6 & 40 18 05 &         &   182.5 &   g& 0729+391  & 07 29 57.1 & 39 11 36 &    65.5 &    65.5 &   Q\\       
   0222+422A & 02 22 22.6 & 42 16 10 &    23.3 &    23.3 &    & 0729+395  & 07 29 36.5 & 39 31 35 &    10.5 &    10.5 &   g\\       
   0224+393  & 02 24 00.9 & 39 18 15 &   185.6 &   185.6 &   Q& 0729+397  & 07 29 33.2 & 39 45 07 &     6.7 &     6.7 &    \\       
   0224+396  & 02 24 52.0 & 39 36 28 &    11.5 &    11.5 &    & 0729+437  & 07 29 11.1 & 43 42 07 &    59.4 &    59.4 &    \\       
   0225+381  & 02 25 25.0 & 38 07 50 &    25.0 &    30.3 &   g& 0730+396  & 07 30 25.0 & 39 41 51 &     9.2 &     9.2 &    \\       
   0225+389  & 02 25 53.8 & 38 55 41 &         &    34.7 &   Q& 0731+438  & 07 31 49.4 & 43 50 57 &    46.7 &    46.7 &   G\\       
   0225+427  & 02 25 44.1 & 42 47 31 &    42.6 &    42.6 &   G& 0733+389  & 07 33 35.8 & 38 59 13 &     6.9 &    11.0 &    \\       
   0226+394  & 02 26 40.6 & 39 29 28 &         &    31.3 &   g& 0735+388  & 07 35 32.7 & 38 53 13 &    17.6 &    28.0 &   g\\       
   0226+396  & 02 26 49.6 & 39 39 47 &     2.5 &     2.5 &    & 0735+390  & 07 35 38.4 & 39 03 05 &    17.9 &    19.0 &   g\\       
   0226+467  & 02 26 05.0 & 46 46 58 &   108.0 &   108.0 &   Q& 0735+395  & 07 35 41.0 & 39 32 37 &    19.6 &    19.6 &    \\       
   0227+397  & 02 27 22.7 & 39 45 27 &     7.9 &     8.1 &    & 0736+386  & 07 36 48.5 & 38 40 45 &    37.5 &    37.5 &    \\       
   0227+398  & 02 27 59.7 & 39 50 15 &    29.8 &    29.8 &    & 0736+398  & 07 36 04.3 & 39 54 04 &    14.2 &    17.3 &    \\       
   0228+392  & 02 28 00.8 & 39 14 05 &    11.9 &    11.9 &    & 0736+400  & 07 36 41.5 & 40 01 47 &    38.8 &    38.8 &   g\\       
   0228+393  & 02 28 41.8 & 39 19 47 &    85.7 &    85.7 &   G& 0739+396  & 07 39 28.4 & 39 36 32 &    10.6 &    11.7 &    \\       
   0228+409A & 02 28 29.6 & 40 56 39 &    47.2 &    47.2 &    & 0739+397A & 07 39 46.0 & 39 46 29 &    18.0 &    19.7 &    \\       
   0231+385  & 02 31 21.6 & 38 31 47 &    84.4 &    84.4 &    & 0739+397B & 07 39 46.0 & 39 48 41 &   251.4 &   251.4 &   Q\\       
   0231+391  & 02 31 45.1 & 39 10 36 &     8.4 &     8.4 &    & 0739+398  & 07 39 13.2 & 39 51 42 &    95.7 &    95.7 &   Q\\       
   0231+405A & 02 31 50.9 & 40 30 52 &    31.2 &    31.2 &    & 0740+380C & 07 40 56.8 & 38 00 31 &    92.2 &    92.2 &   Q\\       
   0232+411B & 02 32 46.2 & 41 10 13 &   156.6 &   156.6 &   Q& 0740+393  & 07 40 35.7 & 39 22 14 &    65.2 &    65.2 &    \\       
   0233+390  & 02 33 43.0 & 39 00 02 &    11.0 &    12.9 &    & 0740+474B & 07 40 48.8 & 47 25 56 &    62.3 &    62.3 &    \\       
   0236+399  & 02 37 00.2 & 39 59 44 &     3.6 &     3.6 &    & 0741+396  & 07 41 07.1 & 39 41 47 &     4.0 &     4.9 &    \\       
   0236+438  & 02 36 52.7 & 43 48 38 &    15.8 &    24.0 &    & 0741+399  & 07 41 57.0 & 39 58 21 &    15.1 &    15.1 &    \\       
   0237+389  & 02 37 42.2 & 38 58 43 &     5.2 &    10.1 &    & 0741+407  & 07 41 52.9 & 40 45 39 &    15.5 &    16.2 &    \\       
   0237+396  & 02 37 46.1 & 39 37 48 &     8.1 &     8.1 &    & 0742+376  & 07 42 20.7 & 37 38 53 &    85.0 &    85.0 &    \\       
   0237+435  & 02 37 14.2 & 43 35 06 &    61.2 &    64.0 &   G& 0742+394  & 07 42 39.5 & 39 24 19 &    12.8 &    13.1 &    \\       
   0239+395  & 02 39 47.8 & 39 31 25 &    25.6 &    26.8 &   g& 0743+392B & 07 43 53.8 & 39 17 19 &    27.1 &    27.1 &    \\       
   0239+397  & 02 39 44.6 & 39 43 02 &    18.3 &    18.3 &   Q& 0743+399  & 07 43 07.0 & 39 58 43 &     1.3 &     1.3 &    \\       
   0240+404  & 02 40 26.8 & 40 28 58 &    32.8 &    32.8 &   g& 0744+399  & 07 44 09.5 & 39 57 38 &    10.2 &    10.2 &    \\       
   0241+393B & 02 41 21.7 & 39 21 20 &         &   221.3 &   g& 0744+464  & 07 44 07.2 & 46 26 25 &    56.1 &    56.1 &   G\\       
   0241+395  & 02 41 29.2 & 39 31 03 &     9.9 &     9.9 &    & 0745+397  & 07 45 38.9 & 39 46 23 &     8.0 &     8.0 &    \\       
   0242+395  & 02 42 18.4 & 39 34 06 &    11.9 &    11.9 &    & 0745+398  & 07 45 18.5 & 39 48 38 &    13.1 &    13.1 &    \\       
   0243+439  & 02 43 02.4 & 43 59 10 &    76.5 &    79.7 &    & 0746+399  & 07 46 59.6 & 39 58 51 &    23.0 &    23.6 &    \\       
   0244+377  & 02 44 23.8 & 37 42 21 &   100.4 &   100.4 &    & 0747+398A & 07 47 02.6 & 39 49 12 &    37.9 &    37.9 &    \\       
   0246+392  & 02 46 22.6 & 39 16 51 &    34.2 &    36.5 &    & 0747+398B & 07 47 57.2 & 39 51 47 &    23.5 &    23.5 &    \\       
   0246+393  & 02 47 04.9 & 39 22 12 &         &   380.8 &   g& 0748+413B & 07 48 54.1 & 41 23 37 &    23.9 &    23.9 &    \\       
   0246+396  & 02 46 48.0 & 39 39 18 &    45.6 &    45.6 &    & 0749+398  & 07 49 15.4 & 39 52 02 &    18.2 &    18.2 &    \\       
   0246+428A & 02 46 15.5 & 42 53 06 &   189.5 &   189.5 &   g& 0749+399  & 07 49 28.5 & 39 54 35 &     9.3 &     9.3 &    \\       
   0247+395  & 02 47 10.0 & 39 30 28 &     5.0 &     5.3 &    & 0750+400  & 07 50 58.1 & 40 03 44 &    10.7 &    15.8 &    \\       
   0247+404  & 02 47 58.2 & 40 29 53 &    46.4 &    46.4 &   G& 0750+402  & 07 50 17.2 & 40 12 40 &     9.2 &    12.1 &    \\       
   0248+392  & 02 48 46.2 & 39 16 08 &    32.7 &    32.7 &    & 0751+392  & 07 51 05.6 & 39 16 04 &    46.1 &    46.1 &    \\       
   0248+396  & 02 48 44.7 & 39 39 13 &     5.8 &     5.8 &    & 0752+398  & 07 52 08.3 & 39 51 18 &     8.5 &     9.5 &    \\       
   0248+467  & 02 47 38.4 & 46 45 03 &         &   379.0 &   g& 0753+383  & 07 53 31.3 & 38 17 51 &    32.2 &    50.0 &    \\       
   0249+383  & 02 49 59.0 & 38 23 11 &   381.6 &   381.6 &   Q& 0753+391  & 07 53 05.8 & 39 09 26 &    19.9 &    35.7 &   g\\       
   0250+384  & 02 50 41.8 & 38 29 27 &    29.9 &    29.9 &    & 0754+394  & 07 54 41.3 & 39 28 50 &     4.4 &     4.8 &   Q\\       
   0250+396  & 02 50 07.7 & 39 41 30 &    26.5 &    28.7 &    & 0754+396  & 07 54 46.9 & 39 37 35 &    48.9 &    48.9 &   G\\       
   0251+393  & 02 51 32.3 & 39 19 23 &   374.8 &   405.9 &   g& 0754+397  & 07 54 36.9 & 39 44 45 &     2.9 &     3.1 &   g\\       
   0252+385  & 02 52 59.8 & 38 30 27 &    10.4 &    11.7 &    & 0755+379B & 07 55 09.0 & 37 55 20 &         &   732.3 &   g\\       
   0252+388  & 02 52 50.0 & 38 52 17 &    14.0 &    14.0 &    & 0756+377  & 07 56 28.4 & 37 47 04 &   168.7 &   172.4 &    \\       
   0252+399  & 02 52 35.0 & 39 54 13 &     5.6 &     5.6 &    & 0756+383  & 07 56 32.9 & 38 22 37 &    62.0 &    64.3 &   g\\       
   0253+396  & 02 53 35.0 & 39 39 52 &     5.4 &     5.4 &    & 0756+406  & 07 56 52.6 & 40 38 11 &    41.9 &    41.9 &   Q\\       
   0254+406  & 02 54 36.6 & 40 38 33 &    63.5 &    63.5 &   G& 0757+395  & 07 57 31.4 & 39 32 54 &         &    27.3 &   g\\       
   0255+460  & 02 55 08.0 & 46 04 08 &    96.7 &    96.7 &   Q& 0757+399  & 07 57 10.6 & 39 55 24 &    13.5 &    13.5 &    \\       
   0258+435  & 02 58 53.9 & 43 31 00 &         &   109.0 &   g& 0759+392  & 07 59 02.8 & 39 12 59 &    11.7 &    11.7 &    \\       
   0258+443  & 02 58 13.7 & 44 18 38 &    68.8 &    68.8 &    & 0759+397  & 07 59 40.9 & 39 45 02 &    15.4 &    15.4 &    \\       
   0259+387  & 02 59 48.5 & 38 44 58 &    21.2 &    25.0 &   g& 0800+399  & 08 00 39.3 & 39 58 35 &    19.8 &    19.8 &   b\\       
   0259+391  & 02 59 18.0 & 39 09 11 &    13.8 &    14.2 &    & 0800+472  & 08 00 38.2 & 47 13 12 &   171.1 &   171.1 &   b\\       
   0700+375  & 07 00 41.1 & 37 31 27 &    51.2 &    51.2 &   G& 0801+394  & 08 01 17.8 & 39 29 06 &     6.3 &     6.3 &    \\       
   0700+390  & 07 00 38.5 & 39 03 00 &    34.9 &    34.9 &    & 0801+399  & 08 01 11.1 & 39 58 28 &         &   <2\phu&    \\       
   0700+398  & 07 00 37.8 & 39 52 55 &     3.1 &     3.1 &    & 0801+401  & 08 01 58.9 & 40 06 18 &    63.6 &    63.6 &    \\       
   0700+399  & 07 00 56.8 & 39 53 51 &    16.2 &    17.1 &    & 0802+398  & 08 02 36.6 & 39 49 21 &    22.5 &    22.5 &   Q\\       
   0701+392  & 07 01 05.1 & 39 15 54 &    77.2 &    77.2 &   Q& 0802+406  & 08 02 08.6 & 40 36 35 &    29.3 &    29.3 &    \\       
   0701+397  & 07 01 44.1 & 39 42 22 &     4.2 &     4.6 &    & 0803+426  & 08 03 16.1 & 42 41 30 &    10.1 &    10.3 &   g\\       
   0701+401  & 07 01 53.4 & 40 07 02 &    76.2 &    79.7 &    & 0803+427  & 08 03 41.0 & 42 43 18 &     3.6 &     5.7 &   g\\       
   0702+396  & 07 02 24.2 & 39 39 25 &     7.8 &     7.8 &    & 0804+399  & 08 04 25.1 & 39 55 23 &     6.1 &     6.1 &   g\\       
   0703+390  & 07 03 22.8 & 39 00 12 &    20.0 &    20.0 &    & 0805+391  & 08 05 02.4 & 39 09 47 &    18.4 &    18.8 &   g\\       
   0703+426A & 07 03 09.6 & 42 36 51 &         &   608.2 &   g& 0805+392  & 08 05 24.4 & 39 15 36 &    10.7 &    10.9 &    \\       
   0703+426B & 07 03 30.6 & 42 37 35 &         &    71.1 &    & 0805+406  & 08 05 40.7 & 40 41 45 &   110.4 &   110.4 &   b\\       
   0703+468  & 07 03 05.9 & 46 52 36 &   280.9 &   280.9 &    & 0806+399  & 08 06 11.4 & 39 58 20 &    37.1 &    37.1 &    \\       
   0704+384  & 07 04 08.3 & 38 26 57 &   106.9 &   110.9 &   Q& 0806+426  & 08 06 37.9 & 42 36 58 &   175.6 &   179.0 &   G\\       
   0704+397  & 07 04 05.8 & 39 46 24 &     4.6 &     4.6 &    & 0807+399  & 08 07 22.1 & 39 59 05 &    13.0 &    13.0 &    \\       
   0704+399  & 07 04 17.8 & 39 54 19 &    14.4 &    14.4 &    & 0809+404  & 08 09 31.6 & 40 28 03 &   194.6 &   194.6 &   g\\       
   0704+418  & 07 04 49.4 & 41 53 39 &    46.9 &    48.6 &    & 0810+460B & 08 10 58.6 & 46 05 48 &   112.8 &   112.8 &   g\\       
   0705+398  & 07 05 52.3 & 39 53 24 &     4.5 &     4.5 &    & 0811+388  & 08 11 53.9 & 38 49 54 &    46.6 &    52.4 &   g\\       
   0706+396  & 07 06 34.5 & 39 41 39 &    13.4 &    13.6 &    & 0811+391  & 08 11 07.9 & 39 08 33 &     9.6 &     9.6 &   G\\       
   0707+380A & 07 07 16.1 & 38 04 14 &    17.6 &    17.6 &    & 0812+382  & 08 12 44.2 & 38 13 29 &    86.2 &    91.2 &   g\\       
   0708+384  & 07 08 25.5 & 38 28 55 &     9.5 &     9.5 &    & 0812+398  & 08 12 38.9 & 39 52 45 &     2.5 &     2.6 &    \\       
   0708+388  & 07 08 43.1 & 38 53 28 &    56.3 &    56.3 &   g& 0812+399  & 08 12 38.1 & 39 58 26 &     5.2 &     5.4 &    \\       
   0709+393  & 07 09 39.9 & 39 18 40 &         &    42.6 &    & 0812+406  & 08 12 46.9 & 40 38 33 &    31.8 &    31.8 &    \\       
   0709+398  & 07 09 22.6 & 39 53 51 &    21.4 &    21.4 &    & 0813+381  & 08 13 34.7 & 38 08 22 &    13.3 &    13.3 &    \\       
   0709+405  & 07 09 32.7 & 40 33 11 &    22.1 &    22.5 &    & 0813+393  & 08 13 34.5 & 39 20 32 &    19.1 &    21.8 &   g\\       
   0709+409  & 07 09 01.1 & 40 56 31 &    38.6 &    41.1 &   g& 0813+398  & 08 13 26.4 & 39 53 26 &    16.9 &    16.9 &   g\\       
   0710+403  & 07 10 56.8 & 40 21 14 &    26.3 &    28.9 &   g& 0814+383  & 08 14 02.0 & 38 20 47 &    17.8 &    17.8 &    \\       
   0710+457  & 07 10 52.3 & 45 45 19 &   247.5 &   247.5 &   g& 0814+425  & 08 14 51.7 & 42 32 08 &  1075.7 &  1075.7 &   BL\\       
   0711+397  & 07 11 51.5 & 39 45 10 &    11.3 &    11.3 &    & 0814+441  & 08 14 08.6 & 44 08 55 &    39.0 &    39.0 &    \\       
   0711+399  & 07 11 05.3 & 39 58 00 &     5.3 &     5.3 &    & 0815+397  & 08 15 25.0 & 39 47 40 &     3.9 &     4.0 &    \\       
   0712+382  & 07 12 26.8 & 38 13 26 &    13.5 &    13.5 &    & 0818+472A & 08 18 00.9 & 47 12 10 &   193.5 &   197.3 &   g\\       
   0712+387  & 07 12 09.8 & 38 44 21 &    23.9 &    30.8 &    & 0819+397  & 08 19 10.5 & 39 45 26 &     7.9 &     7.9 &    \\       
   0712+388  & 07 12 24.5 & 38 53 54 &    15.2 &    15.2 &    & 0820+392  & 08 20 06.1 & 39 16 07 &    23.3 &    25.7 &   g\\       
   0712+390  & 07 12 12.0 & 39 05 23 &    12.6 &    12.6 &    & 0820+431  & 08 20 02.9 & 43 06 38 &   103.1 &   103.1 &   g\\       
   0712+391  & 07 12 03.8 & 39 06 26 &    34.6 &    37.8 &    & 0821+394  & 08 21 37.4 & 39 26 29 &  1741.1 &  1741.1 &   Q\\       
   0714+405  & 07 14 43.4 & 40 30 38 &    10.9 &    13.5 &    & 0821+395  & 08 21 32.0 & 39 33 18 &         &   <5\phu&    \\       
   0717+393  & 07 17 22.3 & 39 21 50 &    11.8 &    12.8 &    & 0821+447  & 08 21 50.3 & 44 46 15 &    89.9 &    96.8 &   Q\\       
   0720+381  & 07 20 43.1 & 38 09 18 &    26.6 &    37.9 &   g& 0822+390  & 08 22 12.6 & 39 02 02 &    15.8 &    15.8 &    \\       
   0720+412  & 07 20 11.5 & 41 14 09 &    17.8 &    18.2 &    & 0822+394  & 08 22 05.5 & 39 29 33 &   126.9 &   126.9 &   G\\       
   0721+394  & 07 21 34.3 & 39 27 18 &    18.8 &    18.8 &    & 0822+398  & 08 22 12.5 & 39 50 54 &     3.8 &     3.8 &    \\       
   0721+398  & 07 21 00.4 & 39 49 48 &     6.3 &     6.3 &    & 0823+384B & 08 23 56.7 & 38 29 53 &    21.2 &    22.4 &   g\\       
   0722+385  & 07 22 21.0 & 38 32 39 &    27.1 &    27.1 &    & 0823+399  & 08 23 35.4 & 39 58 37 &    16.9 &    17.5 &   g\\       
   0722+393A & 07 22 25.1 & 39 23 32 &    59.8 &    59.8 &    & 0824+397  & 08 24 48.8 & 39 45 40 &    14.8 &    14.8 &   Q\\       
   0723+397  & 07 23 28.5 & 39 45 52 &    21.1 &    22.0 &    & 0827+378  & 08 27 55.2 & 37 52 18 &   494.5 &   494.5 &   Q\\       
   0724+396  & 07 24 45.0 & 39 36 33 &    10.5 &    10.5 &   Q& 0827+387  & 08 27 08.8 & 38 47 57 &    15.1 &    15.1 &    \\       
   0726+402  & 07 26 37.9 & 40 17 19 &    47.3 &    47.3 &    & 0827+395  & 08 27 01.6 & 39 33 52 &    12.5 &    12.8 &   g\\\hline       
\end{tabular}
\end{tiny}
\end{flushleft}
\end{table*}

\setcounter{table}{0}
\begin{table*}
\caption[]{B3\,VLA sources at 10.6 GHz (cont'd)}
\begin{flushleft}
\begin{tiny}
\begin{tabular}{|lccrrc|lccrrc|}
\hline
  B3 name & RA (B1950) & DEC (B1950) & S$_{\rm peak}$ & S$_{\rm int}$ & ID &  B3 name & RA (B1950) & DEC (B1950) & S$_{\rm peak}$ & S$_{\rm int}$ & ID\\
 & [$^{\rm h}$ ~ $^{\rm m}$ ~ $^{\rm s}$] & [\degr$\;\;\;$\arcmin$\;\;\;$\arcsec] & [mJy] & [mJy] & & & [$^{\rm h}$ ~ $^{\rm m}$ ~ $^{\rm s}$] & [\degr$\;\;\;$\arcmin$\;\;\;$\arcsec] & [mJy] & [mJy] &\\\hline
   0827+458  & 08 27 07.3 & 45 53 36 &    45.8 &    47.5 &    & 0944+390B & 09 44 23.6 & 39 00 46 &    15.2 &    15.2 &   g\\       
   0828+381  & 08 28 35.8 & 38 06 43 &    17.7 &    18.7 &   g& 0944+397  & 09 44 44.8 & 39 47 07 &     4.6 &     4.6 &    \\       
   0829+395  & 08 29 58.3 & 39 31 34 &    29.5 &    29.5 &    & 0945+408  & 09 45 50.1 & 40 53 43 &  1391.5 &  1391.5 &   Q\\       
   0829+425  & 08 29 26.4 & 42 35 13 &   135.1 &   135.1 &   Q& 0945+419  & 09 45 23.7 & 41 55 41 &    39.7 &    41.0 &    \\       
   0831+393  & 08 31 38.4 & 39 21 13 &    15.4 &    15.4 &    & 0947+405  & 09 47 23.7 & 40 31 59 &    16.4 &    16.4 &    \\       
   0831+399  & 08 31 16.3 & 39 55 13 &         &    16.7 &    & 0947+424  & 09 47 11.8 & 42 27 00 &    60.2 &    60.2 &    \\       
   0832+395  & 08 32 14.6 & 39 33 09 &    36.2 &    36.2 &    & 0948+390  & 09 48 58.6 & 39 04 38 &    15.4 &    15.4 &   g\\       
   0832+399  & 08 32 21.8 & 39 55 04 &     5.4 &     5.7 &    & 0950+402  & 09 50 59.1 & 40 14 02 &    12.0 &    13.1 &    \\       
   0834+399  & 08 34 08.9 & 39 57 53 &     7.8 &     7.8 &    & 0951+398  & 09 51 27.3 & 39 52 11 &    15.0 &    15.0 &   g\\       
   0834+450A & 08 34 27.1 & 45 00 54 &         &   301.3 &   g& 0951+408  & 09 51 39.2 & 40 50 57 &    32.7 &    34.1 &   Q\\       
   0836+399  & 08 36 55.2 & 39 54 29 &     7.8 &     7.8 &    & 0951+422  & 09 51 07.0 & 42 15 19 &   110.3 &   110.3 &   g\\       
   0836+402  & 08 36 54.3 & 40 14 38 &    84.5 &    84.5 &   g& 0953+382  & 09 53 43.4 & 38 17 19 &    28.9 &    28.9 &    \\       
   0836+426  & 08 36 35.6 & 42 38 33 &   207.5 &   207.5 &   Q& 0953+398  & 09 53 05.9 & 39 49 31 &   129.0 &   129.0 &   Q\\       
   0837+399  & 08 37 51.4 & 39 55 30 &     5.7 &     6.2 &    & 0954+436  & 09 54 39.2 & 43 41 08 &    51.0 &    51.0 &   g\\       
   0838+396  & 08 38 54.6 & 39 37 58 &     8.1 &     8.1 &    & 0955+380  & 09 55 30.4 & 38 01 37 &    11.1 &    11.1 &    \\       
   0840+400  & 08 40 50.5 & 40 03 53 &     2.4 &     2.4 &    & 0955+387  & 09 55 01.6 & 38 44 19 &    44.0 &    46.1 &   Q\\       
   0840+424A & 08 40 11.5 & 42 26 20 &   272.0 &   272.0 &    & 0955+390  & 09 55 42.9 & 39 02 44 &    58.9 &    58.9 &    \\       
   0841+386  & 08 41 14.8 & 38 41 49 &   245.1 &   245.1 &    & 0955+396  & 09 55 55.0 & 39 36 52 &    17.9 &    17.9 &    \\       
   0841+397  & 08 41 06.0 & 39 45 03 &         &   <3\phu&    & 0956+391  & 09 56 42.4 & 39 11 26 &    12.3 &    12.3 &    \\       
   0841+403  & 08 41 39.4 & 40 19 09 &    46.5 &    54.6 &   g& 0956+404  & 09 56 34.8 & 40 29 52 &     7.1 &     7.1 &    \\       
   0841+407  & 08 41 51.5 & 40 42 12 &    35.0 &    38.0 &    & 0956+475  & 09 56 08.0 & 47 35 37 &    29.9 &    29.9 &   G\\       
   0842+401  & 08 42 47.6 & 40 07 55 &    21.1 &    22.7 &    & 0957+399  & 09 57 44.4 & 39 55 12 &     4.3 &     4.3 &    \\       
   0843+425  & 08 43 59.0 & 42 34 42 &         &    51.1 &   g& 0958+390A & 09 58 23.8 & 39 02 24 &    15.8 &    15.8 &    \\       
   0844+396  & 08 44 35.5 & 39 41 06 &     7.3 &     7.3 &    & 0958+391  & 09 58 17.7 & 39 06 14 &     2.8 &     2.9 &    \\       
   0847+406  & 08 47 24.8 & 40 40 28 &    12.8 &    12.8 &    & 1004+446  & 10 04 14.3 & 44 39 53 &   141.9 &   168.1 &    \\       
   0849+394  & 08 49 24.9 & 39 30 26 &     5.4 &     6.2 &    & 1007+417  & 10 07 26.1 & 41 47 16 &   323.1 &   335.3 &   Q\\       
   0849+424  & 08 49 15.6 & 42 26 47 &    57.3 &    57.3 &   Q& 1007+422  & 10 07 22.8 & 42 14 19 &    72.7 &    72.7 &    \\       
   0850+383  & 08 50 41.3 & 38 22 39 &    12.0 &    12.0 &    & 1008+395  & 10 08 56.2 & 39 34 54 &     8.3 &     8.3 &    \\       
   0852+384  & 08 52 43.2 & 38 25 03 &    24.0 &    24.4 &    & 1008+423  & 10 08 53.5 & 42 19 23 &   101.4 &   101.4 &    \\       
   0854+399B & 08 54 29.0 & 39 57 12 &    70.9 &    70.9 &   G& 1008+467A & 10 08 39.3 & 46 43 09 &    92.8 &    92.8 &   G\\       
   0855+397  & 08 55 02.1 & 39 42 33 &     3.8 &     4.5 &    & 1009+389A & 10 09 26.0 & 38 58 56 &    16.8 &    16.8 &    \\       
   0855+419  & 08 55 44.0 & 41 54 59 &    25.4 &    25.4 &    & 1009+434  & 10 09 07.0 & 43 27 55 &     9.2 &     9.2 &   G\\       
   0856+397  & 08 56 25.4 & 39 42 04 &     9.6 &    10.2 &    & 1012+389  & 10 12 59.2 & 38 56 50 &    18.9 &    18.9 &    \\       
   0856+406  & 08 56 45.1 & 40 36 18 &    10.4 &    10.4 &   G& 1012+395  & 10 12 52.6 & 39 35 04 &     8.4 &     9.2 &    \\       
   0857+391  & 08 57 41.7 & 39 07 59 &    82.6 &    86.4 &   g& 1012+425  & 10 12 31.5 & 42 34 45 &    22.0 &    23.6 &    \\       
   0857+402  & 08 57 15.9 & 40 16 43 &    22.4 &    22.4 &    & 1013+410  & 10 12 58.2 & 41 01 46 &         &   186.3 &   g\\       
   0858+386  & 08 58 20.8 & 38 38 59 &    58.0 &    58.0 &    & 1014+392  & 10 14 16.5 & 39 16 23 &   228.6 &   228.6 &   g\\       
   0858+388  & 08 58 03.0 & 38 53 55 &    10.5 &    10.5 &    & 1014+397A & 10 14 19.5 & 39 46 32 &         &   110.2 &   g\\       
   0858+452  & 08 58 54.7 & 45 12 43 &   104.2 &   104.2 &    & 1015+383  & 10 15 29.0 & 38 20 30 &    61.5 &    65.9 &   Q\\       
   0859+470  & 08 59 39.9 & 47 02 56 &   933.4 &   933.4 &   Q& 1016+388B & 10 16 50.5 & 38 48 40 &    18.0 &    19.3 &    \\       
   0900+380B & 09 00 49.8 & 38 04 28 &    15.5 &    16.0 &    & 1016+396  & 10 16 29.5 & 39 37 45 &         &   <3\phu&    \\       
   0900+389  & 09 01 00.3 & 38 58 26 &         &    23.5 &    & 1016+397  & 10 16 57.5 & 39 45 25 &     9.4 &     9.4 &   g\\       
   0900+395  & 09 00 13.4 & 39 30 35 &     5.9 &     9.0 &    & 1016+443  & 10 16 46.5 & 44 23 30 &    21.1 &    21.1 &   g\\       
   0900+428  & 09 00 58.8 & 42 50 02 &   450.9 &   450.9 &   g& 1018+393  & 10 18 41.1 & 39 19 03 &    10.3 &    10.3 &    \\       
   0902+383  & 09 02 17.7 & 38 19 26 &    27.6 &    28.3 &    & 1018+405  & 10 18 47.1 & 40 34 48 &    20.2 &    20.7 &    \\       
   0902+384  & 09 02 02.5 & 38 26 34 &    11.0 &    11.0 &    & 1019+382A & 10 19 27.2 & 38 18 02 &    39.5 &    39.5 &    \\       
   0902+414  & 09 02 48.0 & 41 28 31 &   132.7 &   132.7 &   g& 1019+394  & 10 19 58.7 & 39 24 00 &    45.4 &    45.4 &   G\\       
   0902+416  & 09 02 07.2 & 41 40 39 &   103.8 &   103.8 &    & 1019+395  & 10 19 18.8 & 39 32 39 &     8.4 &     8.4 &    \\       
   0903+428  & 09 03 09.7 & 42 51 08 &    26.8 &    26.8 &   G& 1019+397  & 10 19 41.3 & 39 47 01 &    25.3 &    26.6 &   Q\\       
   0904+386  & 09 04 34.9 & 38 39 46 &    27.0 &    27.0 &   Q& 1020+400  & 10 20 14.6 & 40 03 28 &   852.1 &   852.1 &   Q\\       
   0904+396  & 09 04 26.8 & 39 36 36 &     5.6 &     6.0 &    & 1021+384  & 10 21 00.9 & 38 24 00 &    40.9 &    40.9 &    \\       
   0904+399  & 09 04 15.5 & 39 56 26 &    36.1 &    36.1 &   g& 1022+432  & 10 22 30.6 & 43 12 58 &    76.0 &    76.0 &    \\       
   0904+417B & 09 04 18.4 & 41 46 49 &   126.9 &   200.2 &   g& 1023+393  & 10 23 08.6 & 39 20 39 &    16.2 &    16.2 &   g\\       
   0905+380A & 09 05 41.3 & 38 00 29 &   158.4 &   161.2 &   G& 1024+463  & 10 24 13.3 & 46 18 09 &   113.9 &   214.7 &   G\\       
   0905+399  & 09 05 05.8 & 39 55 24 &    16.9 &    16.9 &   G& 1025+390B & 10 25 49.4 & 38 59 57 &   220.3 &   220.3 &   g\\       
   0906+383  & 09 06 53.3 & 38 18 31 &    20.4 &    20.4 &    & 1025+394  & 10 25 55.9 & 39 26 08 &    12.6 &    12.6 &    \\       
   0906+421  & 09 06 30.8 & 42 09 33 &     4.5 &     4.6 &   g& 1027+383  & 10 27 46.3 & 38 18 46 &    32.0 &    32.0 &    \\       
   0906+430  & 09 06 17.3 & 43 05 59 &  1128.4 &  1128.4 &   Q& 1027+390  & 10 27 31.7 & 39 02 25 &    11.3 &    12.1 &    \\       
   0907+381  & 09 07 45.0 & 38 11 32 &   106.9 &   106.9 &   Q& 1027+392  & 10 27 21.7 & 39 13 18 &    80.7 &    80.7 &    \\       
   0908+380B & 09 08 39.6 & 38 02 37 &    26.3 &    26.3 &   G& 1028+400  & 10 28 21.8 & 40 02 21 &    19.8 &    19.8 &    \\       
   0908+380C & 09 08 54.0 & 38 03 54 &   103.2 &   108.7 &   G& 1028+402  & 10 28 07.4 & 40 16 09 &    26.0 &    26.0 &    \\       
   0909+395  & 09 09 30.0 & 39 35 07 &     6.0 &     6.0 &    & 1030+398  & 10 30 27.5 & 39 51 20 &   382.3 &   382.3 &   G\\       
   0909+432  & 09 09 44.9 & 43 17 44 &    36.1 &    36.1 &    & 1030+415  & 10 30 07.8 & 41 31 34 &   230.3 &   230.3 &   Q\\       
   0910+392  & 09 10 41.6 & 39 14 35 &    19.2 &    19.2 &   Q& 1033+387A & 10 33 30.3 & 38 47 07 &         &    21.5 &   g\\       
   0911+384  & 09 11 27.9 & 38 29 10 &    29.7 &    29.7 &    & 1033+387B & 10 33 52.8 & 38 42 19 &     7.9 &     8.1 &    \\       
   0911+395  & 09 11 28.1 & 39 35 08 &    16.0 &    16.0 &    & 1033+388  & 10 33 41.8 & 38 51 07 &         &    31.8 &   g\\       
   0911+418  & 09 11 32.2 & 41 49 36 &   114.2 &   114.2 &   g& 1033+408  & 10 33 31.3 & 40 51 01 &    25.3 &    27.0 &    \\       
   0912+388  & 09 12 25.0 & 38 50 26 &    15.1 &    15.1 &    & 1034+397  & 10 34 40.5 & 39 43 36 &     5.7 &     6.6 &    \\       
   0912+392  & 09 12 55.5 & 39 12 51 &     7.2 &     7.2 &   Q& 1034+404  & 10 34 18.3 & 40 27 13 &    84.0 &    84.0 &   G\\       
   0913+385  & 09 13 39.3 & 38 30 41 &    59.5 &    62.7 &   g& 1035+398  & 10 35 23.5 & 39 48 32 &    16.9 &    16.9 &    \\       
   0913+391  & 09 13 39.5 & 39 07 02 &   487.0 &   487.0 &   Q& 1037+399  & 10 37 19.2 & 39 58 06 &         &     8.6 &    \\       
   0914+390  & 09 14 32.7 & 39 01 54 &     6.9 &     8.7 &    & 1038+398  & 10 38 36.5 & 39 49 38 &    16.5 &    16.5 &    \\       
   0917+458A & 09 17 50.4 & 45 51 47 &         &  1279.0 &   g& 1039+397  & 10 39 08.7 & 39 42 41 &    47.9 &    49.3 &   g\\       
   0918+381  & 09 18 38.8 & 38 06 56 &    90.3 &   113.7 &   Q& 1039+424  & 10 39 11.8 & 42 26 13 &    23.2 &    23.2 &    \\       
   0918+395  & 09 18 24.6 & 39 30 31 &    10.5 &    11.1 &   g& 1040+395  & 10 40 13.2 & 39 30 00 &    50.3 &    50.3 &    \\       
   0918+444  & 09 18 15.8 & 44 26 33 &    84.6 &    84.6 &   g& 1040+397  & 10 40 25.3 & 39 45 22 &     8.5 &     8.5 &    \\       
   0919+381  & 09 19 08.0 & 38 06 52 &    27.9 &    27.9 &   G& 1040+398  & 10 40 43.8 & 39 49 57 &     8.5 &     9.3 &   g\\       
   0920+390  & 09 20 06.3 & 39 02 32 &   299.0 &   299.0 &    & 1041+392  & 10 41 34.5 & 39 16 36 &    30.3 &    30.3 &    \\       
   0920+408  & 09 20 52.4 & 40 47 54 &    50.8 &    53.5 &   g& 1042+392  & 10 42 23.6 & 39 12 25 &    87.8 &    87.8 &    \\       
   0921+400  & 09 21 40.8 & 40 03 27 &    20.7 &    20.7 &    & 1042+393  & 10 42 29.7 & 39 20 06 &     7.3 &     8.4 &    \\       
   0922+397  & 09 22 44.1 & 39 48 06 &     2.4 &     2.4 &    & 1042+397  & 10 42 48.6 & 39 45 56 &    27.1 &    27.1 &    \\       
   0922+407  & 09 22 50.7 & 40 42 49 &   326.6 &   326.6 &   Q& 1043+394  & 10 43 06.4 & 39 28 55 &    13.3 &    14.2 &    \\       
   0922+422  & 09 22 47.8 & 42 16 35 &    20.9 &    20.9 &   Q& 1044+454  & 10 44 38.6 & 45 24 43 &    23.6 &    23.6 &    \\       
   0922+425  & 09 22 11.9 & 42 30 26 &    30.5 &    30.5 &   Q& 1047+387  & 10 47 56.0 & 38 47 43 &    28.5 &    28.5 &    \\       
   0923+392  & 09 23 55.3 & 39 15 24 & 12040.8 & 12040.8 &   Q& 1047+396  & 10 47 23.1 & 39 41 37 &     5.9 &     5.9 &    \\       
   0923+398  & 09 23 38.3 & 39 50 47 &    75.6 &    75.6 &   g& 1049+384  & 10 49 22.3 & 38 27 39 &    71.2 &    71.2 &    \\       
   0926+388  & 09 26 34.5 & 38 49 12 &    10.8 &    10.8 &   Q& 1050+391  & 10 50 44.2 & 39 09 46 &    18.7 &    18.7 &    \\       
   0926+392  & 09 26 08.4 & 39 12 23 &    12.8 &    12.8 &    & 1052+380  & 10 52 56.1 & 38 02 02 &    10.7 &    10.7 &    \\       
   0929+395  & 09 29 18.2 & 39 31 52 &    19.0 &    19.0 &    & 1052+389  & 10 52 36.0 & 38 56 19 &     4.5 &     4.8 &    \\       
   0930+389  & 09 30 00.7 & 38 55 10 &    26.3 &    26.3 &   G& 1053+384  & 10 53 23.9 & 38 24 45 &    42.1 &    52.4 &    \\       
   0930+395  & 09 30 35.0 & 39 35 10 &    25.7 &    26.8 &    & 1053+394  & 10 53 59.8 & 39 27 28 &    15.3 &    16.4 &    \\       
   0931+399  & 09 31 59.7 & 39 55 30 &   190.9 &   190.9 &   g& 1054+396  & 10 54 37.8 & 39 41 18 &    17.0 &    17.0 &    \\       
   0932+397  & 09 32 05.2 & 39 46 26 &    12.0 &    12.0 &    & 1055+381  & 10 55 20.3 & 38 10 19 &    40.5 &    46.6 &   G\\       
   0934+387  & 09 34 48.2 & 38 46 14 &    32.4 &    32.4 &    & 1055+396  & 10 55 51.7 & 39 35 51 &    19.9 &    19.9 &    \\       
   0935+397  & 09 35 33.0 & 39 47 53 &    45.4 &    45.4 &   Q& 1055+404A & 10 55 11.1 & 40 26 23 &    48.8 &    48.8 &    \\       
   0935+428A & 09 35 06.6 & 42 52 07 &    42.4 &    42.4 &   G& 1055+404B & 10 55 48.8 & 40 30 12 &    41.2 &    49.0 &    \\       
   0936+398  & 09 36 17.9 & 39 53 19 &    11.6 &    11.6 &    & 1056+387  & 10 56 28.3 & 38 41 59 &    25.3 &    25.9 &    \\       
   0936+405  & 09 36 13.7 & 40 30 26 &    37.8 &    41.6 &   g& 1056+396  & 10 56 22.8 & 39 41 10 &    22.5 &    22.5 &   G\\       
   0937+391  & 09 37 58.8 & 39 07 34 &    83.5 &   104.2 &   Q& 1056+399  & 10 56 15.0 & 40 00 07 &    24.4 &    24.4 &    \\       
   0937+396  & 09 37 38.7 & 39 42 14 &     5.8 &     5.8 &    & 1056+432A & 10 56 08.2 & 43 17 29 &   329.4 &   329.4 &   G\\       
   0938+399B & 09 38 18.3 & 39 58 23 &         &   341.6 &   g& 1058+393  & 10 58 42.2 & 39 20 40 &   177.2 &   177.2 &    \\       
   0942+394  & 09 42 51.9 & 39 28 49 &     7.8 &     7.8 &    & 1100+398  & 11 00 36.2 & 39 49 56 &     5.9 &     5.9 &    \\       
   0942+399  & 09 42 56.6 & 39 56 43 &    14.9 &    14.9 &    & 1101+384  & 11 01 40.5 & 38 28 43 &   522.4 &   522.4 &   g\\       
   0943+399  & 09 43 13.9 & 39 58 10 &    22.2 &    22.2 &    & 1101+395  & 11 01 50.5 & 39 31 02 &     5.3 &     5.3 &    \\       
   0944+390A & 09 44 07.3 & 39 05 10 &     8.5 &     8.5 &    & 1101+396  & 11 01 15.4 & 39 37 06 &         &   <5\phu&    \\\hline       
 \end{tabular}
\end{tiny}
\end{flushleft}
\end{table*}

\setcounter{table}{0}
\begin{table*}
\caption[]{B3\,VLA sources at 10.6 GHz (cont'd)}
\begin{flushleft}
\begin{tiny}
\begin{tabular}{|lccrrc|lccrrc|}
\hline
  B3 name & RA (B1950) & DEC (B1950) & S$_{\rm peak}$ & S$_{\rm int}$ & ID &  B3 name & RA (B1950) & DEC (B1950) & S$_{\rm peak}$ & S$_{\rm int}$ & ID\\
 & [$^{\rm h}$ ~ $^{\rm m}$ ~ $^{\rm s}$] & [\degr$\;\;\;$\arcmin$\;\;\;$\arcsec] & [mJy] & [mJy] & & & [$^{\rm h}$ ~ $^{\rm m}$ ~ $^{\rm s}$] & [\degr$\;\;\;$\arcmin$\;\;\;$\arcsec] & [mJy] & [mJy] &\\\hline
  1103+393  & 11 03 21.4 & 39 22 17 &    28.0 &    28.0 &    & 1217+427  & 12 17 25.1 & 42 46 29 &    37.4 &    37.4 &    \\       
   1104+390  & 11 04 54.0 & 39 05 18 &    12.8 &    14.6 &    & 1218+395  & 12 18 42.5 & 39 35 24 &     8.9 &    11.2 &    \\       
   1104+397  & 11 04 14.3 & 39 45 27 &         &   <4\phu&    & 1218+398  & 12 18 24.8 & 39 53 50 &     4.3 &     4.3 &    \\       
   1105+390  & 11 05 36.4 & 39 02 07 &    17.1 &    19.3 &   g& 1218+421  & 12 18 44.1 & 42 08 56 &    91.0 &   101.3 &   g\\       
   1105+392  & 11 05 51.8 & 39 14 55 &   128.1 &   128.1 &   Q& 1219+382  & 12 19 42.4 & 38 15 52 &     8.8 &     8.8 &    \\       
   1106+380  & 11 06 43.4 & 38 00 48 &   502.2 &   502.2 &    & 1220+393  & 12 20 54.0 & 39 21 00 &    10.1 &    11.7 &    \\       
   1107+379  & 11 07 04.3 & 37 54 46 &   293.3 &   293.3 &   g& 1220+408  & 12 20 06.8 & 40 52 59 &    47.4 &    47.4 &    \\       
   1108+394  & 11 08 28.7 & 39 28 51 &         &   <5\phu&    & 1221+394  & 12 21 41.7 & 39 25 27 &    11.7 &    11.7 &    \\       
   1108+399  & 11 08 33.7 & 39 56 32 &    79.4 &    79.4 &   g& 1221+397  & 12 21 39.6 & 39 46 33 &    11.3 &    11.3 &    \\       
   1108+411B & 11 08 52.8 & 41 06 35 &    94.6 &    97.4 &   g& 1221+398  & 12 21 19.4 & 39 49 21 &    11.7 &    13.7 &   g\\       
   1109+437  & 11 09 52.0 & 43 42 21 &   144.6 &   144.6 &   Q& 1222+390  & 12 22 53.2 & 39 05 06 &    13.5 &    13.5 &    \\       
   1110+391  & 11 10 41.6 & 39 09 45 &    12.1 &    13.5 &    & 1222+398  & 12 22 03.3 & 39 48 54 &         &   <4\phu&    \\       
   1111+391  & 11 11 44.6 & 39 06 56 &    23.9 &    23.9 &    & 1222+423  & 12 22 00.6 & 42 23 09 &   148.5 &   148.5 &   G\\       
   1111+396A & 11 11 53.0 & 39 43 35 &     5.1 &     5.1 &    & 1223+395  & 12 23 22.6 & 39 30 59 &   331.1 &   331.1 &   G\\       
   1111+408  & 11 11 53.0 & 40 53 41 &   249.1 &   249.1 &   Q& 1225+403  & 12 25 45.9 & 40 20 48 &     8.9 &     8.9 &    \\       
   1112+435  & 11 12 34.4 & 43 31 05 &    63.5 &    63.5 &   g& 1225+442  & 12 25 15.7 & 44 17 16 &    37.4 &    37.4 &   g\\       
   1115+380A & 11 15 21.0 & 38 04 31 &    34.9 &    34.9 &    & 1226+395  & 12 26 44.9 & 39 32 26 &     9.4 &     9.4 &    \\       
   1115+399  & 11 15 12.8 & 39 57 08 &    19.4 &    19.4 &    & 1228+397  & 12 28 26.2 & 39 46 32 &    26.4 &    26.4 &   Q\\       
   1116+388  & 11 16 11.4 & 38 50 50 &    12.8 &    12.8 &    & 1228+419A & 12 28 08.5 & 41 55 28 &         &   154.7 &   g\\       
   1116+392  & 11 16 19.7 & 39 15 17 &    15.6 &    15.6 &   Q& 1229+397  & 12 29 32.5 & 39 47 13 &    16.3 &    16.3 &    \\       
   1117+441  & 11 17 30.7 & 44 11 17 &    54.9 &    63.4 &    & 1229+405  & 12 29 14.2 & 40 34 05 &    58.3 &    58.3 &   Q\\       
   1118+390  & 11 18 29.6 & 39 00 35 &     7.1 &     7.1 &    & 1230+398  & 12 30 17.4 & 39 53 25 &     8.9 &     8.9 &    \\       
   1121+399  & 11 21 04.3 & 39 54 18 &    39.2 &    39.2 &    & 1231+394  & 12 31 57.3 & 39 25 21 &    14.8 &    14.8 &    \\       
   1121+435  & 11 21 48.2 & 43 32 09 &    62.1 &    62.1 &    & 1231+432  & 12 31 53.7 & 43 13 44 &    39.8 &    41.0 &    \\       
   1121+444  & 11 21 10.5 & 44 25 03 &    27.8 &    31.3 &    & 1232+394  & 12 32 45.7 & 39 27 30 &    11.2 &    11.2 &    \\       
   1122+390  & 11 22 01.2 & 39 02 16 &    33.2 &    35.9 &   g& 1232+397A & 12 32 03.9 & 39 47 06 &    20.6 &    21.1 &   G\\       
   1122+397  & 11 23 00.1 & 39 42 01 &    19.5 &    22.3 &    & 1232+397B & 12 32 39.0 & 39 42 10 &    15.7 &    15.7 &   G\\       
   1123+395  & 11 23 46.2 & 39 35 13 &    26.3 &    26.3 &   Q& 1232+399  & 12 32 24.8 & 39 55 15 &    15.7 &    15.7 &    \\       
   1127+380  & 11 27 14.6 & 38 04 37 &    30.6 &    30.6 &    & 1232+414A & 12 32 05.0 & 41 26 05 &    92.9 &    92.9 &   g\\       
   1128+385  & 11 28 12.5 & 38 31 51 &   943.8 &   943.8 &   Q& 1233+399  & 12 33 50.2 & 39 56 33 &     5.2 &     5.2 &    \\       
   1128+392  & 11 28 31.8 & 39 17 07 &    20.6 &    20.6 &    & 1233+418  & 12 33 10.8 & 41 53 38 &   131.1 &   131.1 &   g\\       
   1128+396  & 11 28 48.3 & 39 38 06 &     8.0 &     8.7 &    & 1234+396  & 12 34 26.2 & 39 36 58 &   170.6 &   170.6 &    \\       
   1128+436  & 11 28 05.4 & 43 41 37 &    38.4 &    38.4 &    & 1236+444A & 12 36 10.5 & 44 30 12 &         &    88.8 &    \\       
   1128+455  & 11 28 56.4 & 45 31 23 &   231.3 &   231.3 &   g& 1236+444B & 12 36 24.1 & 44 26 15 &         &    38.7 &   g\\       
   1130+387  & 11 30 17.0 & 38 43 31 &    28.3 &    28.3 &    & 1239+382  & 12 39 47.8 & 38 15 22 &    29.4 &    34.6 &   g\\       
   1131+388  & 11 31 23.7 & 38 52 24 &    52.1 &    62.6 &   g& 1239+390  & 12 39 09.1 & 39 05 06 &    26.9 &    28.2 &    \\       
   1131+391  & 11 31 33.1 & 39 07 52 &    72.4 &    74.3 &    & 1239+396  & 12 39 23.8 & 39 37 14 &    35.9 &    35.9 &   g\\       
   1131+437  & 11 31 57.3 & 43 44 36 &   215.7 &   215.7 &   G& 1239+442B & 12 39 57.3 & 44 12 37 &    71.5 &    91.0 &   Q\\       
   1132+396  & 11 32 22.0 & 39 39 37 &     4.0 &     4.0 &    & 1240+381  & 12 40 27.0 & 38 07 25 &   340.7 &   340.7 &   Q\\       
   1132+406  & 11 32 00.6 & 40 37 51 &    37.6 &    41.1 &   g& 1240+395  & 12 40 29.0 & 39 32 11 &     7.7 &     7.7 &    \\       
   1132+410  & 11 32 05.3 & 41 00 28 &    29.1 &    29.1 &    & 1241+411  & 12 41 57.1 & 41 08 00 &    73.5 &    73.5 &   g\\       
   1133+395  & 11 33 46.3 & 39 33 51 &    12.8 &    12.8 &    & 1242+391  & 12 42 06.7 & 39 10 21 &     5.3 &     6.8 &    \\       
   1133+432  & 11 33 15.3 & 43 15 20 &   172.3 &   172.3 &    & 1242+410  & 12 42 26.4 & 41 04 29 &   339.9 &   339.9 &   Q\\       
   1134+406  & 11 34 47.1 & 40 39 07 &    29.8 &    29.8 &    & 1244+389  & 12 44 22.9 & 38 58 02 &    48.6 &    52.1 &    \\       
   1135+390  & 11 35 18.8 & 39 01 45 &     7.5 &     7.5 &    & 1244+397  & 12 44 57.0 & 39 44 10 &     6.7 &     6.7 &    \\       
   1135+401  & 11 35 01.1 & 40 07 25 &    33.7 &    33.7 &    & 1245+389  & 12 45 57.6 & 38 58 36 &    32.2 &    32.9 &    \\       
   1136+383  & 11 36 55.2 & 38 20 19 &    84.0 &    84.0 &    & 1245+396  & 12 45 41.0 & 39 38 26 &     9.4 &    11.3 &    \\       
   1136+390  & 11 36 30.5 & 39 03 55 &    32.2 &    32.2 &    & 1245+399  & 12 45 26.9 & 39 56 45 &    13.7 &    14.4 &    \\       
   1136+420  & 11 36 19.5 & 42 05 17 &    75.6 &    75.6 &   G& 1246+385C & 12 46 50.8 & 38 33 16 &    24.5 &    38.3 &    \\       
   1137+396  & 11 37 46.2 & 39 38 38 &     6.5 &     7.2 &   g& 1247+450A & 12 47 03.3 & 45 01 09 &    92.8 &    97.0 &   Q\\       
   1140+394  & 11 40 36.4 & 39 25 50 &     7.1 &     7.8 &    & 1249+393  & 12 49 23.7 & 39 19 28 &    30.7 &    30.7 &    \\       
   1140+399  & 11 40 41.3 & 39 55 05 &     6.9 &     7.8 &    & 1249+432  & 12 49 41.0 & 43 14 00 &    69.0 &    69.0 &    \\       
   1141+374  & 11 41 49.4 & 37 25 19 &   442.0 &   442.0 &   g& 1249+475  & 12 49 58.8 & 47 31 55 &    68.5 &    88.4 &    \\       
   1141+392  & 11 41 10.9 & 39 15 21 &    29.1 &    29.1 &    & 1250+384  & 12 50 09.6 & 38 27 00 &    21.1 &    21.1 &   g\\       
   1141+400  & 11 41 55.1 & 40 00 02 &    27.1 &    27.1 &   Q& 1250+390  & 12 50 27.8 & 39 05 53 &    14.1 &    14.1 &    \\       
   1141+466  & 11 41 00.4 & 46 37 59 &    60.3 &    60.3 &   g& 1251+398  & 12 51 49.4 & 39 49 38 &     6.5 &     7.3 &   Q\\       
   1142+392  & 11 42 56.5 & 39 13 36 &    32.0 &    32.0 &   Q& 1253+374  & 12 53 55.9 & 37 29 54 &    59.8 &    73.4 &    \\       
   1143+405  & 11 43 50.0 & 40 31 38 &    12.7 &    12.7 &    & 1253+432  & 12 53 24.5 & 43 14 39 &    58.8 &    58.8 &   G\\       
   1143+456  & 11 43 37.0 & 45 37 17 &    43.3 &    43.3 &   G& 1254+476  & 12 54 40.9 & 47 36 32 &   727.0 &   727.0 &   G\\       
   1144+398  & 11 44 40.0 & 39 53 32 &     8.6 &     8.6 &    & 1255+448  & 12 55 43.9 & 44 51 33 &   133.7 &   149.2 &    \\       
   1144+402  & 11 44 21.0 & 40 15 13 &   918.4 &   918.4 &   Q& 1256+392  & 12 56 42.2 & 39 16 30 &    42.8 &    42.8 &   Q\\       
   1144+404  & 11 44 05.5 & 40 24 41 &    34.8 &    34.8 &    & 1257+383  & 12 57 53.3 & 38 20 39 &    58.7 &    67.9 &    \\       
   1148+387  & 11 48 53.3 & 38 42 33 &   139.5 &   139.5 &   Q& 1257+399  & 12 57 25.7 & 39 56 46 &     3.6 &     4.1 &    \\       
   1148+477  & 11 48 32.3 & 47 45 36 &    52.2 &    52.2 &   Q& 1258+395  & 12 58 24.0 & 39 35 12 &     9.2 &    12.1 &    \\       
   1149+390  & 11 49 01.8 & 39 02 55 &     7.0 &     7.0 &    & 1258+404  & 12 58 13.8 & 40 25 16 &   134.3 &   139.9 &   Q\\       
   1149+398  & 11 49 08.6 & 39 50 55 &    15.3 &    15.3 &    & 1259+395  & 12 59 41.2 & 39 31 29 &    14.5 &    15.9 &   Q\\       
   1150+388  & 11 50 09.4 & 38 48 26 &    33.5 &    33.5 &    & 1300+397  & 13 00 29.1 & 39 46 00 &    21.5 &    21.5 &   Q\\       
   1150+401  & 11 50 25.4 & 40 10 33 &    14.6 &    17.1 &    & 1301+382  & 13 01 24.3 & 38 12 14 &    74.5 &    79.9 &   g\\       
   1150+438  & 11 50 06.2 & 43 53 00 &    31.2 &    31.2 &    & 1301+393  & 13 01 01.5 & 39 19 23 &    12.1 &    13.3 &    \\       
   1151+383  & 11 51 25.9 & 38 21 49 &    67.7 &    67.7 &   g& 1302+388  & 13 02 38.8 & 38 48 32 &    31.7 &    31.7 &   G\\       
   1151+384B & 11 51 17.3 & 38 28 27 &   128.1 &   222.2 &   g& 1305+393  & 13 05 19.0 & 39 20 43 &    17.3 &    17.3 &    \\       
   1151+408  & 11 51 19.0 & 40 53 34 &   396.6 &   396.6 &   Q& 1306+396  & 13 06 10.5 & 39 42 01 &     5.2 &     5.5 &    \\       
   1151+456  & 11 51 44.8 & 45 40 19 &   203.3 &   203.3 &   g& 1308+392  & 13 08 20.6 & 39 12 57 &    36.3 &    36.3 &    \\       
   1153+407B & 11 53 17.6 & 40 47 26 &    29.6 &    29.6 &    & 1309+412A & 13 09 27.1 & 41 14 53 &         &   116.9 &   g\\       
   1153+451  & 11 53 33.6 & 45 06 57 &    93.1 &   105.7 &    & 1311+419  & 13 11 49.5 & 41 56 27 &    27.9 &    33.1 &    \\       
   1154+397  & 11 54 12.7 & 39 45 00 &         &    16.1 &   g& 1312+393  & 13 12 49.5 & 39 19 34 &    39.7 &    39.7 &   Q\\       
   1154+398  & 11 54 54.8 & 39 52 48 &     5.0 &     5.0 &    & 1313+387  & 13 13 07.4 & 38 46 20 &         &    45.8 &    \\       
   1156+389  & 11 56 30.2 & 38 56 51 &    16.7 &    17.9 &   G& 1313+392  & 13 13 25.2 & 39 14 22 &    28.8 &    28.8 &   g\\       
   1157+396  & 11 57 17.3 & 39 41 00 &     3.1 &     3.1 &    & 1314+453A & 13 13 59.8 & 45 20 18 &    93.5 &    93.5 &   G\\       
   1157+460  & 11 57 57.2 & 46 05 24 &   121.3 &   121.3 &   G& 1315+395  & 13 15 34.5 & 39 31 05 &    30.3 &    30.8 &    \\       
   1158+393  & 11 58 21.8 & 39 19 39 &     9.7 &    10.2 &    & 1315+396  & 13 15 02.8 & 39 41 14 &   285.2 &   285.2 &   Q\\       
   1159+395  & 11 59 16.3 & 39 35 52 &   116.7 &   116.7 &   G& 1317+380  & 13 17 36.2 & 38 03 07 &    46.6 &    46.6 &   Q\\       
   1200+393  & 12 00 26.9 & 39 22 28 &     8.1 &     8.1 &    & 1317+389  & 13 17 45.0 & 38 56 05 &    80.9 &    80.9 &    \\       
   1201+394  & 12 01 33.7 & 39 29 00 &    74.5 &    74.5 &   g& 1317+393  & 13 17 28.0 & 39 18 35 &    14.1 &    14.5 &   G\\       
   1201+396  & 12 01 09.4 & 39 39 37 &     5.8 &     5.8 &    & 1318+398  & 13 18 12.6 & 39 48 36 &     9.9 &     9.9 &    \\       
   1202+388  & 12 02 58.4 & 38 50 45 &    23.5 &    23.5 &    & 1319+388  & 13 19 04.7 & 38 51 13 &    14.6 &    14.6 &    \\       
   1202+397  & 12 02 15.9 & 39 46 29 &     7.2 &     7.2 &    & 1319+397  & 13 19 57.6 & 39 46 29 &     8.5 &     8.5 &    \\       
   1203+384  & 12 03 44.5 & 38 29 15 &    32.0 &    37.0 &   Q& 1319+398  & 13 19 33.9 & 39 53 55 &    13.2 &    13.2 &    \\       
   1204+399  & 12 04 04.6 & 39 57 45 &   280.2 &   280.2 &   Q& 1318+428C & 13 19 05.4 & 42 50 47 &         &   418.2 &   g\\       
   1204+401  & 12 04 33.9 & 40 11 21 &    22.0 &    22.0 &   G& 1320+389  & 13 20 06.4 & 38 59 44 &     9.6 &    10.1 &    \\       
   1205+390  & 12 05 25.2 & 39 05 31 &     5.9 &     5.9 &    & 1321+415  & 13 21 11.0 & 41 30 52 &    88.8 &    88.8 &   g\\       
   1205+392  & 12 05 20.2 & 39 12 40 &    73.4 &    79.6 &   g& 1322+398  & 13 22 30.9 & 39 49 19 &     5.9 &     5.9 &    \\       
   1206+399A & 12 06 31.6 & 39 56 57 &     9.0 &     9.5 &   g& 1324+390  & 13 24 47.8 & 39 05 23 &    11.4 &    11.4 &    \\       
   1206+399B & 12 06 36.7 & 39 55 24 &     9.0 &     9.5 &   g& 1324+431  & 13 24 51.1 & 43 10 03 &    25.1 &    26.4 &    \\       
   1206+439B & 12 06 42.0 & 43 56 00 &   262.0 &   262.0 &   Q& 1327+390  & 13 27 51.5 & 39 00 55 &    10.7 &    10.7 &    \\       
   1207+386  & 12 07 37.6 & 38 37 55 &    36.0 &    38.4 &   G& 1327+391  & 13 27 15.2 & 39 09 43 &     6.7 &     6.7 &    \\       
   1208+396  & 12 08 01.0 & 39 41 02 &    69.3 &    69.3 &   g& 1327+398  & 13 27 54.1 & 39 52 46 &     4.2 &     6.8 &    \\       
   1209+396  & 12 09 56.7 & 39 39 41 &    30.9 &    31.8 &    & 1327+474C & 13 27 46.3 & 47 27 09 &   235.0 &   235.0 &   g\\       
   1209+399  & 12 09 24.3 & 39 59 30 &     7.2 &     7.2 &    & 1328+388  & 13 28 58.2 & 38 50 26 &    33.4 &    33.4 &    \\       
   1211+388  & 12 11 59.5 & 38 51 49 &     9.2 &     9.2 &    & 1328+396  & 13 28 45.7 & 39 36 43 &    58.3 &    58.3 &    \\       
   1212+380  & 12 12 26.1 & 38 05 31 &    22.9 &    22.9 &    & 1330+380  & 13 30 36.4 & 38 00 36 &         &    31.7 &    \\       
   1212+389  & 12 12 43.2 & 38 56 03 &    19.6 &    19.6 &   Q& 1330+389  & 13 30 14.1 & 38 55 33 &     6.4 &     6.5 &    \\       
   1213+389  & 12 13 53.2 & 38 54 57 &    13.6 &    13.9 &    & 1330+406  & 13 30 44.1 & 40 39 46 &    18.6 &    19.3 &    \\       
   1216+400  & 12 16 34.8 & 40 04 30 &    14.3 &    14.3 &    & 1331+381  & 13 31 15.7 & 38 11 21 &    16.5 &    17.2 &    \\       
   1216+402  & 12 16 07.5 & 40 17 25 &    50.8 &    50.8 &   G& 1332+385  & 13 32 12.5 & 38 33 17 &    47.6 &    54.1 &   g\\ \hline      
 \end{tabular}
\end{tiny}
\end{flushleft}
\end{table*}
\clearpage

\setcounter{table}{0}
\begin{table*}
\caption[]{B3\,VLA sources at 10.6 GHz (cont'd)}
\begin{flushleft}
\begin{tiny}
\begin{tabular}{|lccrrc|lccrrc|}
\hline
  B3 name & RA (B1950) & DEC (B1950) & S$_{\rm peak}$ & S$_{\rm int}$ & ID &  B3 name & RA (B1950) & DEC (B1950) & S$_{\rm peak}$ & S$_{\rm int}$ & ID\\
 & [$^{\rm h}$ ~ $^{\rm m}$ ~ $^{\rm s}$] & [\degr$\;\;\;$\arcmin$\;\;\;$\arcsec] & [mJy] & [mJy] & & & [$^{\rm h}$ ~ $^{\rm m}$ ~ $^{\rm s}$] & [\degr$\;\;\;$\arcmin$\;\;\;$\arcsec] & [mJy] & [mJy] &\\\hline
   1333+392  & 13 33 38.7 & 39 15 13 &    15.2 &    15.2 &    & 1446+399  & 14 46 10.8 & 39 56 47 &    10.7 &    10.7 &    \\       
   1333+412  & 13 33 09.5 & 41 15 22 &    94.6 &    94.6 &   g& 1446+440  & 14 46 41.3 & 44 05 01 &    70.0 &    70.0 &    \\       
   1334+417  & 13 34 16.6 & 41 46 29 &    62.3 &    62.3 &   G& 1447+380  & 14 47 18.1 & 38 01 06 &    10.0 &    10.0 &    \\       
   1336+391A & 13 36 38.3 & 39 06 24 &   362.6 &   362.6 &   g& 1447+400  & 14 47 33.3 & 40 00 46 &    40.0 &    50.2 &    \\       
   1336+393  & 13 36 04.0 & 39 21 10 &    16.2 &    16.2 &    & 1447+402  & 14 47 06.2 & 40 12 45 &         &   104.4 &   g\\       
   1336+396C & 13 36 57.8 & 39 41 34 &    50.7 &    58.0 &    & 1449+380  & 14 49 01.2 & 38 03 50 &    17.0 &    17.4 &    \\       
   1336+397A & 13 36 05.3 & 39 43 45 &    47.5 &    47.5 &    & 1449+421  & 14 49 14.2 & 42 07 00 &    40.6 &    40.6 &    \\       
   1336+397B & 13 36 44.2 & 39 45 07 &     3.2 &     3.2 &    & 1450+391B & 14 50 08.9 & 39 10 34 &         &    71.7 &   g\\       
   1337+385  & 13 37 13.3 & 38 30 20 &     8.1 &     8.1 &    & 1450+396  & 14 50 07.4 & 39 38 07 &     7.4 &     7.4 &    \\       
   1338+394  & 13 38 57.8 & 39 29 59 &    15.1 &    15.5 &   Q& 1451+396  & 14 51 53.3 & 39 37 59 &     7.9 &     7.9 &    \\       
   1339+438  & 13 39 46.8 & 43 50 26 &    37.3 &    39.5 &   G& 1452+394  & 14 52 59.4 & 39 23 58 &    23.6 &    23.6 &   g\\       
   1339+472  & 13 39 41.8 & 47 12 23 &    85.8 &    85.8 &   Q& 1453+397  & 14 53 34.5 & 39 46 03 &     9.1 &     9.1 &    \\       
   1340+439  & 13 40 59.9 & 43 58 27 &    40.9 &    40.9 &    & 1454+394  & 14 54 24.2 & 39 26 48 &    12.4 &    12.4 &   G\\       
   1341+392  & 13 41 10.1 & 39 13 39 &    69.1 &    71.3 &   Q& 1455+399  & 14 55 52.1 & 39 57 58 &     2.5 &     2.5 &    \\       
   1342+389A & 13 42 14.8 & 38 56 31 &    44.2 &    44.2 &   Q& 1455+421  & 14 55 49.0 & 42 10 48 &   114.7 &   114.7 &   G\\       
   1342+389B & 13 42 37.9 & 38 59 37 &    24.6 &    25.1 &   g& 1457+388A & 14 57 14.8 & 38 48 31 &    11.2 &    12.1 &    \\       
   1343+386  & 13 43 26.6 & 38 38 12 &   194.7 &   194.7 &   Q& 1458+433  & 14 58 40.5 & 43 21 41 &    61.8 &    61.8 &   G\\       
   1343+430  & 13 43 26.7 & 43 05 16 &   160.4 &   160.4 &   G& 1459+399  & 14 59 15.0 & 39 54 29 &    18.6 &    18.6 &    \\       
   1344+397  & 13 44 55.2 & 39 43 55 &    77.1 &    77.1 &    & 2300+382  & 23 00 54.4 & 38 12 28 &    33.5 &    33.5 &   g\\       
   1345+398  & 13 45 04.5 & 39 50 54 &     6.7 &     6.7 &    & 2301+394A & 23 01 28.4 & 39 25 13 &    25.0 &    25.0 &    \\       
   1346+392  & 13 46 47.8 & 39 15 36 &    13.5 &    13.5 &    & 2301+394B & 23 01 36.7 & 39 26 56 &    19.3 &    19.3 &    \\       
   1347+391  & 13 47 16.6 & 39 06 34 &    32.0 &    32.0 &    & 2301+398  & 23 01 06.9 & 39 50 04 &     8.2 &     9.2 &   G\\       
   1347+396  & 13 47 00.5 & 39 37 59 &    28.9 &    28.9 &    & 2301+430  & 23 01 29.0 & 43 00 49 &    36.1 &    36.1 &    \\       
   1347+398  & 13 47 07.4 & 39 51 39 &    20.3 &    21.6 &    & 2301+443  & 23 01 27.8 & 44 22 55 &    60.9 &    60.9 &    \\       
   1347+403  & 13 47 44.2 & 40 21 12 &    12.9 &    12.9 &    & 2302+396  & 23 02 27.2 & 39 40 32 &         &   <3\phu&    \\       
   1348+392  & 13 48 23.6 & 39 14 12 &   101.4 &   101.4 &   Q& 2302+402  & 23 02 34.8 & 40 12 41 &   159.8 &   159.8 &    \\       
   1348+396  & 13 48 31.4 & 39 36 15 &    26.6 &    26.6 &    & 2303+391A & 23 03 44.0 & 39 10 55 &         &   151.1 &   g\\       
   1349+394  & 13 49 41.3 & 39 28 31 &    12.4 &    12.4 &    & 2304+377  & 23 04 39.4 & 37 46 28 &   234.0 &   234.0 &   g\\       
   1349+388  & 13 49 10.9 & 38 53 02 &   115.7 &   115.7 &   Q& 2304+398  & 23 04 42.6 & 39 53 45 &    12.1 &    12.6 &    \\       
   1350+395  & 13 50 21.3 & 39 33 22 &     7.3 &     7.3 &    & 2304+429  & 23 04 13.0 & 42 54 33 &    36.4 &    36.4 &    \\       
   1350+432  & 13 50 24.0 & 43 14 09 &    10.1 &    10.1 &   G& 2305+404  & 23 05 33.1 & 40 25 34 &    16.2 &    16.2 &   G\\       
   1352+383  & 13 52 54.5 & 38 19 35 &    16.3 &    18.4 &    & 2306+392  & 23 06 34.6 & 39 17 19 &    10.7 &    10.7 &    \\       
   1352+397  & 13 52 03.7 & 39 42 38 &    16.3 &    16.3 &    & 2308+393  & 23 08 35.4 & 39 22 21 &    15.0 &    15.0 &    \\       
   1352+403  & 13 52 00.3 & 40 20 39 &    22.8 &    24.7 &    & 2308+395  & 23 08 19.4 & 39 35 24 &     7.0 &     7.0 &    \\       
   1353+380  & 13 53 44.9 & 38 00 03 &    21.8 &    24.2 &    & 2308+400  & 23 08 32.7 & 40 03 51 &    16.9 &    16.9 &    \\       
   1353+397  & 13 53 56.3 & 39 48 01 &    14.3 &    14.3 &    & 2311+387  & 23 11 37.5 & 38 45 29 &    31.7 &    31.7 &    \\       
   1354+397  & 13 54 08.8 & 39 44 02 &         &    11.7 &    & 2311+396A & 23 11 27.9 & 39 36 42 &         &   <3\phu&   F\\       
   1355+380  & 13 55 31.5 & 38 04 17 &    60.6 &    71.3 &   Q& 2311+396B & 23 11 53.8 & 39 38 51 &     3.7 &     3.7 &    \\       
   1356+393  & 13 56 40.8 & 39 18 35 &    41.3 &    41.3 &    & 2311+469  & 23 11 29.2 & 46 55 55 &   243.6 &   243.6 &   Q\\       
   1356+397  & 13 56 08.8 & 39 47 21 &     7.4 &     8.7 &   Q& 2313+406  & 23 13 47.2 & 40 38 54 &    25.5 &    25.5 &    \\       
   1357+392  & 13 57 05.3 & 39 15 05 &     6.2 &     6.7 &   b& 2315+396  & 23 15 25.1 & 39 36 13 &   149.5 &   211.7 &   g\\       
   1357+394A & 13 57 48.8 & 39 27 34 &    14.2 &    14.2 &    & 2316+398  & 23 16 57.3 & 39 53 43 &    20.1 &    22.1 &   Q\\       
   1357+394B & 13 57 56.5 & 39 25 23 &    21.4 &    22.7 &   Q& 2318+389  & 23 18 54.8 & 38 55 48 &     9.9 &     9.9 &    \\       
   1358+433  & 13 58 29.6 & 43 18 32 &    37.3 &    42.4 &    & 2320+396  & 23 20 19.9 & 39 41 33 &    25.3 &    27.1 &    \\       
   1359+419  & 13 59 30.0 & 41 56 42 &    49.3 &    49.3 &   G& 2320+416B & 23 20 21.5 & 41 42 11 &         &    68.1 &   g\\       
   1401+387  & 14 01 05.1 & 38 42 14 &   114.5 &   114.5 &    & 2321+423  & 23 21 30.3 & 42 19 19 &    96.4 &    96.4 &   G\\       
   1401+395  & 14 01 59.3 & 39 35 30 &     5.9 &     5.9 &    & 2322+396  & 23 22 52.8 & 39 41 06 &   116.7 &   116.7 &   F\\       
   1402+382  & 14 02 21.1 & 38 14 49 &    18.2 &    20.5 &    & 2322+403  & 23 22 23.8 & 40 23 52 &    33.0 &    33.0 &    \\       
   1403+395  & 14 03 06.6 & 39 31 41 &     8.0 &     8.2 &    & 2323+388  & 23 23 14.8 & 38 49 25 &    30.9 &    30.9 &    \\       
   1406+397  & 14 06 20.4 & 39 45 42 &    18.3 &    18.3 &    & 2323+398  & 23 23 06.2 & 39 51 49 &     4.7 &     4.9 &    \\       
   1407+388  & 14 07 21.7 & 38 52 10 &    23.3 &    23.3 &   g& 2323+435A & 23 23 18.4 & 43 30 28 &   368.1 &   368.1 &   g\\       
   1408+398  & 14 08 37.3 & 39 48 42 &    15.5 &    16.1 &    & 2324+394B & 23 24 42.6 & 39 26 48 &     9.6 &     9.8 &    \\       
   1408+399  & 14 08 13.7 & 39 59 32 &     4.0 &     4.2 &    & 2324+405  & 23 24 30.5 & 40 31 36 &   425.8 &   441.5 &    \\       
   1409+390  & 14 09 37.7 & 39 00 23 &    28.6 &    28.6 &    & 2325+396  & 23 25 33.2 & 39 39 10 &    15.0 &    15.0 &    \\       
   1409+394  & 14 09 22.7 & 39 28 21 &    13.9 &    14.6 &    & 2326+395  & 23 26 24.9 & 39 31 25 &    22.3 &    22.3 &    \\       
   1410+438  & 14 10 18.3 & 43 51 54 &    55.9 &    55.9 &    & 2326+422  & 23 26 33.0 & 42 15 31 &    92.7 &    96.2 &   g\\       
   1411+391A & 14 11 48.8 & 39 07 26 &    18.9 &    18.9 &    & 2327+391  & 23 27 58.5 & 39 10 39 &    14.4 &    14.4 &   G\\       
   1411+391B & 14 11 56.1 & 39 09 24 &    30.9 &    37.9 &    & 2327+407  & 23 27 42.7 & 40 47 52 &    82.7 &    82.7 &   Q\\       
   1411+397  & 14 11 09.3 & 39 44 56 &    13.9 &    14.2 &   g& 2327+422  & 23 27 39.5 & 42 17 13 &    44.9 &    45.7 &    \\       
   1411+427  & 14 11 40.7 & 42 43 22 &    32.7 &    37.8 &    & 2328+397  & 23 29 00.0 & 39 46 58 &     5.3 &     5.3 &    \\       
   1412+392  & 14 12 59.5 & 39 13 11 &    53.5 &    53.5 &    & 2329+398  & 23 29 47.0 & 39 50 45 &    10.1 &    11.9 &   Q\\       
   1412+397  & 14 12 46.7 & 39 44 01 &     5.5 &     5.5 &    & 2330+387  & 23 30 35.4 & 38 44 38 &   251.5 &   251.5 &   g\\       
   1413+398  & 14 13 12.8 & 39 51 37 &     3.9 &     3.9 &    & 2330+389  & 23 30 01.3 & 38 58 28 &    10.5 &    10.5 &    \\       
   1414+398  & 14 14 30.0 & 39 50 26 &    11.1 &    11.1 &   g& 2330+397  & 23 30 27.2 & 39 45 21 &    12.8 &    12.8 &    \\       
   1415+391  & 14 15 51.2 & 39 11 52 &    10.2 &    10.2 &    & 2330+402  & 23 30 26.2 & 40 14 02 &   149.8 &   149.8 &    \\       
   1416+400  & 14 16 56.7 & 40 00 25 &    68.2 &    74.1 &   Q& 2330+435  & 23 30 56.3 & 43 30 12 &    15.6 &    15.6 &   g\\       
   1417+383  & 14 17 51.6 & 38 20 09 &    30.3 &    30.3 &    & 2331+399  & 23 31 17.2 & 39 55 53 &    41.0 &    42.0 &   G\\       
   1417+385  & 14 17 43.0 & 38 35 32 &   899.2 &   899.2 &   Q& 2332+388  & 23 32 34.8 & 38 49 34 &    19.0 &    19.0 &   Q\\       
   1417+397  & 14 17 46.6 & 39 46 04 &     5.0 &     7.4 &    & 2332+399  & 23 32 07.1 & 39 56 47 &    22.7 &    22.7 &    \\       
   1418+388  & 14 18 58.6 & 38 49 25 &    10.2 &    10.2 &    & 2333+397  & 23 33 01.4 & 39 43 33 &     4.3 &     4.4 &    \\       
   1419+397  & 14 19 21.6 & 39 47 11 &    97.6 &    97.6 &   g& 2333+399  & 23 33 06.2 & 39 55 07 &    10.6 &    10.6 &    \\       
   1419+399  & 14 19 23.9 & 39 57 08 &    51.5 &    56.5 &   Q& 2334+398  & 23 34 27.9 & 39 49 09 &    43.9 &    45.0 &   g\\       
   1419+419  & 14 19 06.4 & 41 58 30 &   372.2 &   372.2 &   g& 2335+392  & 23 35 43.0 & 39 16 56 &    32.4 &    32.4 &   Q\\       
   1420+386  & 14 20 10.6 & 38 40 32 &    23.3 &    26.3 &    & 2336+381  & 23 36 24.9 & 38 06 47 &    39.4 &    41.9 &    \\       
   1422+395  & 14 22 20.8 & 39 35 16 &         &    33.4 &   g& 2337+398  & 23 37 15.0 & 39 52 07 &    26.3 &    26.3 &   g\\       
   1422+401B & 14 22 27.9 & 40 06 51 &    44.9 &    44.9 &    & 2338+390  & 23 38 44.9 & 39 01 51 &    40.5 &    40.5 &   Q\\       
   1424+380  & 14 24 02.9 & 38 02 26 &         &   <3\phu&    & 2338+393  & 23 38 52.0 & 39 19 26 &    35.4 &    35.4 &    \\       
   1426+394  & 14 26 12.9 & 39 25 41 &   155.3 &   155.3 &    & 2340+386  & 23 40 00.2 & 38 38 07 &    18.3 &    20.3 &    \\       
   1426+398  & 14 26 45.3 & 39 51 59 &     6.2 &     6.2 &    & 2341+396A & 23 41 37.5 & 39 37 15 &         &    20.3 &   g\\       
   1427+404  & 14 27 57.4 & 40 24 35 &    23.2 &    28.9 &   g& 2341+396B & 23 41 44.2 & 39 35 26 &         &    22.6 &   g\\       
   1428+380  & 14 28 32.3 & 38 03 26 &    31.5 &    40.0 &   g& 2341+399  & 23 41 54.5 & 39 56 12 &    33.8 &    34.4 &    \\       
   1428+385  & 14 28 53.7 & 38 32 51 &    28.6 &    36.3 &    & 2342+394  & 23 42 29.2 & 39 29 46 &     2.3 &     2.3 &    \\       
   1429+392  & 14 29 36.1 & 39 12 48 &     7.5 &     7.5 &    & 2344+429  & 23 44 52.9 & 42 54 12 &   254.2 &   254.2 &   Q\\       
   1429+395  & 14 29 42.3 & 39 32 18 &    19.2 &    19.2 &    & 2347+397  & 23 47 54.1 & 39 45 21 &     4.2 &     4.2 &    \\       
   1430+399  & 14 30 27.5 & 39 58 01 &     3.5 &     6.6 &    & 2348+387  & 23 48 03.1 & 38 47 40 &    14.1 &    21.2 &    \\       
   1432+382  & 14 32 56.4 & 38 17 54 &    69.0 &    71.8 &   g& 2348+450  & 23 48 57.0 & 45 01 49 &    87.9 &    87.9 &   G\\       
   1432+389  & 14 32 14.0 & 38 56 03 &     8.8 &     8.8 &    & 2349+396  & 23 49 31.8 & 39 38 21 &    21.8 &    21.8 &    \\       
   1432+397A & 14 32 19.8 & 39 42 39 &         &   <3\phu&    & 2349+410  & 23 49 21.5 & 41 04 33 &    47.6 &    47.6 &   Q\\       
   1432+397B & 14 32 57.1 & 39 47 13 &    16.1 &    16.1 &   g& 2350+395  & 23 50 17.2 & 39 31 13 &   262.0 &   262.0 &    \\       
   1432+428B & 14 32 32.4 & 42 49 25 &   182.8 &   182.8 &    & 2351+394  & 23 51 09.3 & 39 26 41 &    16.9 &    20.1 &    \\       
   1435+383  & 14 35 32.9 & 38 20 42 &    47.5 &    47.5 &   Q& 2351+398  & 23 51 09.1 & 39 49 00 &     2.4 &     2.4 &    \\       
   1435+429  & 14 34 59.0 & 42 57 15 &    25.6 &    26.2 &    & 2351+400B & 23 51 25.2 & 40 01 13 &         &   157.8 &    \\       
   1436+399  & 14 36 41.2 & 39 55 53 &     7.3 &     7.3 &    & 2351+456  & 23 51 50.0 & 45 36 22 &   727.3 &   727.3 &   Q\\       
   1437+397  & 14 37 12.1 & 39 46 28 &    16.4 &    16.4 &    & 2352+385  & 23 52 05.8 & 38 30 53 &    33.0 &    33.0 &    \\       
   1437+427  & 14 37 51.8 & 42 47 18 &    56.5 &    56.5 &   G& 2354+397  & 23 54 34.5 & 39 46 20 &    10.1 &    10.1 &    \\       
   1438+382  & 14 38 11.5 & 38 14 46 &    14.4 &    14.4 &    & 2354+471  & 23 54 58.0 & 47 09 24 &         &   383.6 &   g\\       
   1438+385  & 14 38 22.4 & 38 33 05 &   640.2 &   640.2 &   Q& 2355+397  & 23 55 38.7 & 39 45 11 &    12.9 &    12.9 &    \\       
   1438+406  & 14 38 30.9 & 40 41 17 &    18.5 &    18.5 &   g& 2355+398  & 23 55 28.0 & 39 49 47 &    27.0 &    27.0 &    \\       
   1441+409  & 14 41 03.1 & 40 57 10 &   143.6 &   143.6 &    & 2356+390  & 23 56 26.5 & 39 05 47 &   241.2 &   241.2 &   Q\\       
   1442+383  & 14 42 17.8 & 38 21 12 &    26.5 &    26.5 &    & 2356+437  & 23 56 03.0 & 43 48 06 &   177.6 &   186.1 &   G\\       
   1442+384  & 14 42 40.4 & 38 29 54 &    23.0 &    23.0 &    & 2357+398  & 23 57 05.3 & 39 49 54 &    10.6 &    10.6 &   g\\       
   1442+441  & 14 42 58.1 & 44 10 18 &    27.6 &    38.5 &    & 2358+390  & 23 58 07.9 & 39 01 22 &   110.4 &   110.4 &    \\       
   1444+395  & 14 44 30.1 & 39 33 32 &    10.2 &    10.2 &    & 2358+406  & 23 58 19.4 & 40 37 19 &   260.1 &   260.1 &    \\       
   1444+417A & 14 44 32.7 & 41 45 51 &    71.5 &    71.5 &   Q& 2358+416  & 23 58 45.8 & 41 36 25 &    29.1 &    31.4 &   G\\       
   1445+410  & 14 45 17.3 & 41 00 15 &   101.6 &   101.6 &   g& 2359+394  & 23 59 09.3 & 39 29 39 &    32.1 &    42.9 &    \\   \hline    \end{tabular}
\end{tiny}
\end{flushleft}
\end{table*}

%% file: ds1562t3.tab.tex
\begin{table*}
\caption[]{B3--VLA Decomposed Double Sources}
%\begin{flushleft}
\begin{tiny}
\begin{tabular}{|lccrr|lccrr|}
\hline
%   & & & & & & & & &  \\
  B3 name & RA (B1950) & DEC (B1950)& F(2.8cm ) & F(20cm)&B3 Name & RA (B1950) & DEC (B1950)& F(2.8cm ) & F(20cm)\\
 & [\degr$\;\;\;$\arcmin$\;\;\;$\arcsec] & [$^{\rm h}$ ~ $^{\rm m}$ ~ $^{\rm s}$]&  [mJy]&   [mJy]&  &[\degr$\;\;\;$\arcmin$\;\;\;$\arcsec] & [$^{\rm h}$ ~ $^{\rm m}$ ~ $^{\rm s}$] &   [mJy]  &   [mJy]\\\hline
 0010+402 & 00 10 14.3 & 40 15 52 &     34.1 &    150 & 1016+397 & 10 17 00.0 & 39 44 43 &      4.0 &     23 \\
          & 00 10 18.6 & 40 16 22 &     15.5 &    150 &          & 10 16 55.0 & 39 46 07 &      5.4 &     28 \\
 0015+399 & 00 15 03.5 & 40 00 26 &     24.7 &     13 & 1027+383 & 10 27 43.7 & 38 18 58 &     13.6 &     55 \\
          & 00 15 03.9 & 39 59 13 &      7.3 &     48 &          & 10 27 48.8 & 38 18 34 &     18.4 &    182 \\
 0017+432 & 00 17 26.4 & 43 13 29 &     14.5 &    162 & 1028+402 & 10 28 07.2 & 40 16 36 &     14.6 &    137 \\
          & 00 17 32.3 & 43 13 33 &     12.1 &     84 &          & 10 28 07.6 & 40 15 43 &     11.4 &     48 \\
 0020+437 & 00 20 49.7 & 43 46 13 &     30.0 &    233 & 1040+395 & 10 40 10.5 & 39 29 39 &     25.9 &     94 \\
          & 00 20 53.0 & 43 46 39 &     14.2 &     99 &          & 10 40 15.9 & 39 30 22 &     24.4 &    119 \\
 0146+394 & 01 46 46.9 & 39 25 26 &      9.3 &     85 & 1047+396 & 10 47 20.8 & 39 41 43 &      3.4 &     26 \\
          & 01 46 47.2 & 39 23 14 &      5.8 &     32 &          & 10 47 25.4 & 39 41 31 &      2.5 &     14 \\
 0152+435 & 01 52 25.3 & 43 30 58 &    150.8 &    987 & 1055+396 & 10 55 45.8 & 39 36 54 &     10.8 &     42 \\
          & 01 52 27.2 & 43 31 43 &    100.8 &    522 &          & 10 55 57.6 & 39 34 48 &      9.1 &     26 \\
 0153+417 & 01 53 17.6 & 41 48 10 &     63.6 &    243 & 1105+392 & 11 05 49.8 & 39 15 16 &    109.8 &    669 \\
          & 01 53 19.6 & 41 47 28 &     36.9 &    320 &          & 11 05 53.9 & 39 14 35 &     18.3 &     90 \\
 0159+390 & 01 59 49.9 & 39 02 26 &     10.3 &     34 & 1107+379 & 11 07 02.6 & 37 54 35 &    148.0 &    975 \\
          & 01 59 53.3 & 39 02 04 &      9.8 &     28 &          & 11 07 06.1 & 37 54 57 &    145.3 &   1090 \\
 0213+392 & 02 13 11.1 & 39 16 56 &      7.5 &     42 & 1109+437 & 11 09 51.2 & 43 42 47 &     48.9 &    554 \\
          & 02 13 06.9 & 39 17 05 &      6.7 &     47 &          & 11 09 52.8 & 43 41 56 &     95.7 &    634 \\
 0216+423 & 02 15 58.8 & 42 19 04 &     79.8 &    429 & 1118+390 & 11 18 28.2 & 39 01 17 &      2.5 &     23 \\
          & 02 16 03.3 & 42 19 23 &     37.4 &    191 &          & 11 18 31.0 & 38 59 54 &      4.6 &     31 \\
 0217+417 & 02 17 04.5 & 41 43 34 &     33.1 &     97 & 1128+436 & 11 28 04.9 & 43 41 10 &     30.9 &    271 \\
          & 02 17 06.5 & 41 43 50 &     21.9 &     32 &          & 11 28 05.9 & 43 42 05 &      7.5 &     64 \\
 0218+396 & 02 18 44.9 & 39 41 18 &     86.4 &    404 & 1141+392 & 11 41 09.3 & 39 15 06 &     15.2 &     89 \\
          & 02 18 40.4 & 39 42 28 &     70.5 &    342 &          & 11 41 12.4 & 39 15 37 &     13.9 &     71 \\
 0218+399A& 02 18 44.9 & 39 55 48 &     10.0 &    100 & 1149+398 & 11 49 04.7 & 39 51 25 &      7.5 &     40 \\
          & 02 18 36.5 & 39 55 01 &     10.4 &    101 &          & 11 49 12.5 & 39 50 28 &      7.8 &     30 \\
 0219+443 & 02 19 05.1 & 44 19 57 &     25.4 &    112 & 1222+423 & 12 21 59.6 & 42 22 45 &     60.6 &    506 \\
          & 02 19 07.2 & 44 19 39 &     38.7 &     88 &          & 12 22 01.6 & 42 23 33 &     87.9 &    636 \\
 0221+383 & 02 21 15.6 & 38 18 19 &      9.8 &     38 & 1230+398 & 12 30 16.0 & 39 53 06 &      3.1 &     26 \\
          & 02 21 08.8 & 38 18 49 &     19.0 &     94 &          & 12 30 18.7 & 39 53 45 &      5.8 &     13 \\
 0231+391 & 02 31 43.3 & 39 10 49 &      4.1 &     33 & 1324+390 & 13 24 48.3 & 39 05 56 &      4.8 &     25 \\
          & 02 31 46.9 & 39 10 23 &      4.3 &     37 &          & 13 24 47.4 & 39 04 50 &      6.6 &     45 \\
 0232+411B& 02 32 41.6 & 41 09 55 &     93.3 &    430 & 1346+392 & 13 46 47.6 & 39 15 00 &      5.9 &     32 \\
          & 02 32 50.8 & 41 10 31 &     63.3 &    350 &          & 13 46 47.9 & 39 16 12 &      7.6 &     37 \\
 0240+404 & 02 40 22.0 & 40 29 12 &     22.5 &    138 & 1348+396 & 13 48 30.4 & 39 36 35 &      9.2 &     40 \\
          & 02 40 31.6 & 40 28 45 &     10.3 &     30 &          & 13 48 32.4 & 39 35 56 &     17.4 &     56 \\
 0250+384 & 02 50 41.2 & 38 29 50 &      6.8 &    159 & 1349+394 & 13 49 39.6 & 39 26 43 &      7.1 &     21 \\
          & 02 50 42.3 & 38 29 04 &     23.1 &    180 &          & 13 49 42.9 & 39 30 20 &      5.3 &     16 \\
 0700+390 & 07 00 44.7 & 39 04 06 &     10.4 &     70 & 1352+397 & 13 52 00.2 & 39 42 02 &     13.9 &     66 \\
          & 07 00 32.2 & 39 01 55 &     24.5 &    100 &          & 13 52 07.2 & 39 43 14 &      2.4 &     22 \\
 0710+457 & 07 10 49.3 & 45 45 27 &    175.1 &    573 & 1401+387 & 14 01 04.6 & 38 42 50 &     73.3 &    290 \\
          & 07 10 55.3 & 45 45 12 &     72.4 &    300 &          & 14 01 05.5 & 38 41 38 &     41.2 &    262 \\
 0742+376 & 07 42 18.8 & 37 39 13 &     42.1 &    427 & 1410+438 & 14 10 20.2 & 43 51 24 &     22.5 &    130 \\
          & 07 42 22.6 & 37 38 33 &     42.9 &     71 &          & 14 10 16.4 & 43 52 25 &     33.4 &    208 \\
 0759+392 & 07 59 02.5 & 39 12 38 &      6.0 &     29 & 1414+398 & 14 14 30.8 & 39 51 35 &      7.6 &     46 \\
          & 07 59 03.2 & 39 13 21 &      5.7 &     37 &          & 14 14 29.2 & 39 49 17 &      3.5 &     15 \\
 0811+391 & 08 11 04.6 & 39 09 29 &      6.9 &     21 & 1415+391 & 14 15 49.9 & 39 12 11 &      2.8 &     11 \\
          & 08 11 11.2 & 39 07 37 &      2.7 &     53 &          & 14 15 52.4 & 39 11 33 &      7.4 &     50 \\
 0813+398 & 08 13 29.6 & 39 53 46 &      7.1 &     45 & 1436+399 & 14 36 36.9 & 39 55 56 &      2.5 &     11 \\
          & 08 13 23.1 & 39 53 06 &      9.8 &     70 &          & 14 36 45.5 & 39 55 50 &      4.8 &     23 \\
 0824+397 & 08 24 47.0 & 39 45 20 &     12.6 &     60 & 1437+427 & 14 37 50.8 & 42 46 53 &     39.7 &    330 \\
          & 08 24 50.7 & 39 46 00 &      2.2 &     20 &          & 14 37 52.8 & 42 47 43 &     16.8 &     94 \\
 0836+402 & 08 36 54.1 & 40 15 03 &     28.7 &    140 & 1442+384 & 14 42 43.5 & 38 29 52 &     10.4 &     53 \\
          & 08 36 54.4 & 40 14 14 &     55.8 &    395 &          & 14 42 37.3 & 38 29 57 &     12.6 &     85 \\
 0850+383 & 08 50 42.1 & 38 23 06 &      4.4 &     32 & 1444+417A& 14 44 28.3 & 41 45 46 &     33.3 &    200 \\
          & 08 50 40.6 & 38 22 13 &      7.6 &     67 &          & 14 44 37.0 & 41 45 57 &     38.2 &    225 \\
 0854+399B& 08 54 34.0 & 39 56 18 &     48.3 &    234 & 1446+399 & 14 46 05.4 & 39 55 26 &      3.4 &     20 \\
          & 08 54 23.9 & 39 58 07 &     22.6 &    100 &          & 14 46 16.1 & 39 58 08 &      7.3 &     50 \\
 0905+399 & 09 05 01.3 & 39 55 37 &     11.2 &    181 & 1446+440 & 14 46 40.1 & 44 04 39 &     43.3 &    260 \\
          & 09 05 10.2 & 39 55 11 &      5.7 &     50 &          & 14 46 42.4 & 44 05 24 &     26.7 &    193 \\
 0910+392 & 09 10 39.5 & 39 14 36 &      8.4 &     43 & 1459+399 & 14 59 08.1 & 39 53 49 &     11.4 &     38 \\
          & 09 10 43.7 & 39 14 35 &     10.8 &     60 &          & 14 59 21.9 & 39 55 09 &      7.2 &     58 \\
 0911+418 & 09 11 31.7 & 41 50 09 &     55.7 &    226 & 2304+429 & 23 04 12.1 & 42 54 15 &     25.5 &    138 \\
          & 09 11 32.7 & 41 49 03 &     58.5 &    24  &          & 23 04 13.8 & 42 54 52 &     10.9 &    116 \\
 0918+444 & 09 18 22.1 & 44 26 53 &     68.0 &    224 & 2308+393 & 23 08 34.5 & 39 22 41 &      8.2 &     36 \\
          & 09 18 09.5 & 44 26 13 &     16.6 &     40 &          & 23 08 36.3 & 39 22 01 &      6.8 &     41 \\
 0934+387 & 09 34 48.1 & 38 45 20 &     22.7 &    129 & 2313+406 & 23 13 46.9 & 40 39 24 &     16.3 &    127 \\
          & 09 34 48.2 & 38 47 08 &      9.7 &     65 &          & 23 13 47.5 & 40 38 25 &      9.2 &     66 \\
 0947+424 & 09 47 09.3 & 42 26 54 &     38.3 &    251 & 2330+435 & 23 30 54.5 & 43 30 08 &      4.6 &    146 \\
          & 09 47 14.2 & 42 27 07 &     21.9 &    166 &          & 23 30 58.2 & 43 30 17 &     11.0 &     88 \\
 0951+398 & 09 51 27.0 & 39 52 38 &     12.3 &     40 & 2338+393 & 23 38 47.6 & 39 20 14 &     16.2 &     74 \\
          & 09 51 27.5 & 39 51 44 &      2.7 &     33 &          & 23 38 56.4 & 39 18 39 &     19.2 &     59 \\
 0956+391 & 09 56 41.0 & 39 12 02 &      5.6 &     29 & 2354+397 & 23 54 32.7 & 39 46 53 &      5.5 &     35 \\
          & 09 56 43.8 & 39 10 51 &      6.7 &     29 &          & 23 54 36.2 & 39 45 47 &      4.6 &     14 \\\hline
\end{tabular}
\end{tiny}
%\end{flushleft}
\end{table*}

%% file: ds1562t4.tab.tex
\begin{table*}[t]
\caption[]{B3\,VLA sources at 10.6 GHz:  the maps}
%\begin{flushleft}
\begin{tiny}
\begin{tabular}{|lcccrrrcrrc|}
\hline
 B3 name      &  & RA (B1950) & DEC (B1950)   & Total flux & Comp. flux & \multicolumn{3}{c}{Dimensions}  & \multicolumn{1}{c}{PA}  & Noise\\
           &  & [$^{\rm h}$ ~ $^{\rm m}$ ~ $^{\rm s}$]& [\degr$\;\;\;$\arcmin$\;\;\;$\arcsec] & [mJy] & [mJy] & [\arcsec & \hspace{-0.3cm}$\times$&\hspace{-0.35cm} \arcsec] & \multicolumn{1}{c}{[$^{\circ}$]}& [mJy/beam]\\\hline
0000+394  & N & 00 00 08.9 &  39 31 23.7 &  17.5\phast &  6.5\phast  & \multicolumn{3}{c}{Point-like}         &     &  0.8 \\
          & S & 00 00 09.2 &  39 28 26.0 &             & 11.0\phast  & \multicolumn{3}{c}{Point-like}         &     &      \\

0005+383B &   & 00 05 47.4 &  38 20 27.4 &  92.5\phast &             & 60.2& \hspace{-0.3cm}$\times$&\hspace{-0.35cm}33.1  &   39&  0.8 \\

0035+385A &   & 00 35 01.5 &  38 31 34.1 &             &  55.6\phast & \multicolumn{3}{c}{Point-like}         &     &      \\
0035+385B &   & 00 35 08.2 &  38 31 14.9 & 126.9\phast &  71.1\phast & 22.2& \hspace{-0.3cm}$\times$&\hspace{-0.35cm}\ph14.2 &   110 &   0.7\\

0050+401  &   & 00 50 45.2 &  40 10 54.8 &             &  73.8\phast & 41.8& \hspace{-0.3cm}$\times$&\hspace{-0.35cm}29.9  &    5   &   \\
0050+402B &   & 00 50 45.4 &  40 14 58.3 & 117.4\phast &  14.2\phast & \multicolumn{3}{c}{Point-like}&&1.2\\
0050+403  &   & 00 50 45.5 &  40 18 13.6 &             &  29.4\phast & 39.5& \hspace{-0.3cm}$\times$&\hspace{-0.35cm}24.0  &  165  &  \\

0115+469  & S & 01 15 23.6 &  46 55 39.3 &             &  32.3\phast & 48.1 &\hspace{-0.3cm}$\times$&\hspace{-0.35cm}17.1  &   23&\\
          & N & 01 15 27.2 &  46 57 14.6 &  64.3\phast &  32.0\phast & \multicolumn{3}{c}{Point-like} &       &  0.5\\

0131+390  & S & 01 31 48.2 &  39 03 45.0 &             &  16.1\phast &     22.9 &\hspace{-0.3cm}$\times$&\hspace{-0.35cm}13.1  &   29&\\
          & N & 01 31 50.5 &  39 05 17.5 &  27.0\phast &  10.9\phast & \multicolumn{3}{c}{Point-like}  &       &  0.3\\

0132+376A &   & 01 32 29.5&   37 38 29.8 &             &  97.0\phast &     43.9 &\hspace{-0.3cm}$\times$&\hspace{-0.35cm}37.8  &   63  &  \\
0132+376B &  & 01 32 39.7 &  37 39 19.6 & 196.7\phast &       99.7\phast   &     52.5 &\hspace{-0.3cm}$\times$&\hspace{-0.35cm}41.3  &   47  &   1.7\\

0157+393B & S & 01 57 47.5&  39 19 00.0 &        &       55.4\phast   &    124.4 &\hspace{-0.3cm}$\times$&\hspace{-0.35cm}57.6 & 41 & \\
          & N & 01 57 51.5 &  39 21 01.2 &  163.5\phast  &      108.1\phast   &     77.2 &\hspace{-0.3cm}$\times$&\hspace{-0.35cm}35.4  &   1 & 0.9\\

0157+405A &  & 01 57 01.6 &  40 36 04.5 & & 49.0$^\ast$&    \multicolumn{3}{c}{---}  &&\\
0157+405B &   & 01 57 49.0&   40 33 01.0 &  94.6$^\ast$&45.6$^\ast$ &        \multicolumn{3}{c}{---}     &         &  1.1\\

0211+393  & W & 02 10 57.6 &  39 18 44.9 &            & 14.9\phast   & \multicolumn{3}{c}{Point-like}  &     &  \\
          & E & 02 11 07.3 &  39 19 09.5 & 28.2\phast & 13.3\phast   & \multicolumn{3}{c}{Point-like}  &     &    0.4\\

0214+393  &   & 02 14 34.0 &  39 22 36.2 & 85.5\phast &              &  70.3 &\hspace{-0.3cm}$\times$&\hspace{-0.35cm}39.9   &  37  &  0.7\\

0219+428A &   & 02 19 29.8 &  42 48 31.9 & 1235.2\phast & & 51.0 &\hspace{-0.3cm}$\times$&\hspace{-0.35cm}\ph19.0 & 175 & 3.0\\

0220+427A &   & 02 20 02.4 &  42 45 50.9 & 1806.3$^\ast$ & &   \multicolumn{3}{c}{---} & & \\

0221+393  & S & 02 21 34.3 &  39 17 57.7 &   &  2.3\phast  &    \multicolumn{3}{c}{Point-like}&&\\
          & N & 02 21 37.9 &  39 20 20.2 & 7.6\phast  &  5.3\phast   &     70.8& \hspace{-0.3cm}$\times$&\hspace{-0.35cm}25.4  &  104 &    0.5\\

0222+403  &   & 02 22 37.8 &  40 18 08.7 & 182.5\phast  &             &     110.3 &\hspace{-0.3cm}$\times$&\hspace{-0.35cm}39.9  & 168  &  1.3\\

0225+389  &   & 02 25 53.2 &  38 54 55.3 & 34.7$^\ast$ & &   \multicolumn{3}{c}{---} & & 1.5\\

0226+394  &   & 02 26 40.9 &  39 29 57.5 & 31.3$^\ast$& &    \multicolumn{3}{c}{---}   &   &  1.1\\

0232+411B &   & 02 32 42.0 &  41 09 59.6 &       &       115.6\phast   &     41.5 &\hspace{-0.3cm}$\times$&\hspace{-0.35cm}28.2 &   75&\\
0232+411C &   & 02 32 50.3 &  41 10 32.7 &  190.1\phast &        74.5\phast   &     35.3 &\hspace{-0.3cm}$\times$&\hspace{-0.35cm}21.7 &   74  &    1.0\\

0241+393B & W & 02 41 20.1 &  39 21 02.5 &   & 156.7\phast & 63.6 &\hspace{-0.3cm}$\times$&\hspace{-0.35cm}26.5 & 62 & \\
          & E & 02 41 27.0 &  39 22 06.3 & 221.3\phast & 64.6\phast & 39.1 &\hspace{-0.3cm}$\times$&\hspace{-0.35cm}21.4 & 122 & 0.8\\

0246+393  & W & 02 46 59.8 &  39 22 19.5 &       &       207.5\phast   &     59.7 &\hspace{-0.3cm}$\times$&\hspace{-0.35cm}40.9 &  102&\\
          & E & 02 47 10.3 &  39 22 05.6 &  380.8\phast &       173.3\phast   &     58.7 &\hspace{-0.3cm}$\times$&\hspace{-0.35cm}40.4 &   85  &   1.2\\

0248+467  &   & 02 47 38.9 &  46 44 52.2 & 379.0$^\ast$ & &   \multicolumn{3}{c}{---} & & 1.1\\

0258+435  &   & 02 58 52.4 &  43 30 48.0 & 109.0\phast & &  \multicolumn{3}{c}{Point-like}&&0.9\\

          & N & 07 03 05.2 &  42 38 33.7 & &  75.2\phast &69.2 &\hspace{-0.3cm}$\times$&\hspace{-0.35cm}48.0 & 127 &\\
0703+426A & C & 07 03 10.0 &  42 36 39.7 &   608.2\phast & 299.8\phast &55.2 &\hspace{-0.3cm}$\times$&\hspace{-0.35cm}40.5 & 148 &0.9\\
          & S & 07 03 12.0 &  42 35 08.3 & & 233.2\phast &45.2 &\hspace{-0.3cm}$\times$&\hspace{-0.35cm}36.2 &  20 &\\

0703+426B & N & 07 03 30.0 &  42 38 12.4 &  71.1\phast & 26.7\phast &\multicolumn{3}{c}{Point-like} & & 0.9\\
          & S & 07 03 30.2 &  42 36 39.7 &             & 44.4\phast & 43.1 &\hspace{-0.3cm}$\times$&\hspace{-0.35cm}22.8 &160\phast &\\

0709+393  & W & 07 09 35.6 &  39 18 51.3 &             & 25.9\phast &  30.0 &\hspace{-0.3cm}$\times$&\hspace{-0.35cm}18.7   &  3 &\\
          & E & 07 09 43.9 &  39 18 17.1 &  42.6\phast & 16.7\phast &\multicolumn{3}{c}{Point-like}            &    &   0.7\\

0755+379B &   & 07 55 09.4 &  37 55 17.4 & 732.3\phast &               &     89.1 &\hspace{-0.3cm}$\times$&\hspace{-0.35cm}39.8  & 109   & 1.9\\

0757+395  &   & 07 57 32.1 &  39 33 05.5 & 27.3$^\ast$ & &   \multicolumn{3}{c}{---} & &  0.7\\    
                                 
0831+399  & W & 08 31 13.0 &  39 55 31.3 &        &       11.1\phast  &      \multicolumn{3}{c}{Point-like}&&\\
          & E & 08 31 21.0 &  39 54 56.4 &  16.7\phast &         5.6\phast  &      \multicolumn{3}{c}{Point-like} &      &   0.9\\

0834+450A & S & 08 34 26.4 &  45 00 05.6 &        &      129.8\phast   &     50.6 &\hspace{-0.3cm}$\times$&\hspace{-0.35cm}29.3 & 130&\\
          & N & 08 34 27.7 &  45 01 27.7 &  301.3\phast  &      171.5\phast  &      53.6 &\hspace{-0.3cm}$\times$&\hspace{-0.35cm}31.7 & 175  &    1.2 \\

0843+425  & S & 08 43 57.5 &  42 34 07.5 &         &      20.4\phast   &     \multicolumn{3}{c}{Point-like}&&\\
          & N & 08 43 59.8 &  42 35 26.3 &   51.1\phast  &       30.7\phast   &     36.8 &\hspace{-0.3cm}$\times$&\hspace{-0.35cm}13.3  & 13   &   1.2\\

0900+389  & N & 09 00 56.9 &  38 59 04.1 &   23.5\phast  &       16.6\phast   &     39.4 &\hspace{-0.3cm}$\times$&\hspace{-0.35cm}22.4  &137  &    0.8\\
          & S & 09 01 01.5 &  38 57 57.4 &       &         6.9\phast   &     \multicolumn{3}{c}{Point-like}  &    &   \\

0917+458A & W & 09 17 47.0 &  45 51 02.2 &          &    659.4\phast   &     63.8 &\hspace{-0.3cm}$\times$&\hspace{-0.35cm}39.2  & 43&\\
          & E & 09 17 53.9 &  45 52 44.5 &  1279.0\phast   &    619.6\phast   &     58.9 &\hspace{-0.3cm}$\times$&\hspace{-0.35cm}46.3  & 44  &     1.9\\

0938+399B &   & 09 38 17.9 &  39 58 25.0 & 341.6\phast &  & 97.5 &\hspace{-0.3cm}$\times$&\hspace{-0.35cm}39.7 & 178 & 2.8\\

1013+410  & W & 10 12 53.0 &  41 01 43.1 &        &       83.8\phast   &     51.0 &\hspace{-0.3cm}$\times$&\hspace{-0.35cm}38.9 &  88&\\
          & E & 10 13 02.9 &  41 02 02.3 &   186.3\phast &      102.5\phast   &     59.0 &\hspace{-0.3cm}$\times$&\hspace{-0.35cm}44.4 &  89   &   1.5\\

1014+397A & N & 10 14 15.4 &  39 47 29.8 &   110.2\phast &       63.9\phast   &    121.2 &\hspace{-0.3cm}$\times$&\hspace{-0.35cm}43.4 & 132  &    1.4\\
          & S & 10 14 22.8 &  39 46 04.6 &        &       46.3\phast   &     81.3 &\hspace{-0.3cm}$\times$&\hspace{-0.35cm}40.9 & 111&\\

1033+388  & S & 10 33 41.9 &  38 50 14.2 &      & 16.8\phast & \multicolumn{3}{c}{Point-like} & & \\
          & N & 10 33 43.2 &  38 51 18.6 & 31.8\phast & 15.0\phast & \multicolumn{3}{c}{Point-like} & &  0.8\\

1037+399  &   & 10 37 19.3 &  39 58 07.3 &  8.6\phast &      & \multicolumn{3}{c}{Point-like} & & 0.8\\

1154+397  &   & 11 54 12.7 &  39 44 52.1 &  16.1\phast &             &     109.5 &\hspace{-0.3cm}$\times$&\hspace{-0.35cm}36.2 & 145  &   0.8\\

1228+419A &   & 12 28 08.3 &  41 55 39.8 &  154.7\phast  &            &     140.5 &\hspace{-0.3cm}$\times$&\hspace{-0.35cm}69.7 & 119  &   1.3\\

1236+444A & N & 12 36 10.6 &  44 30 20.8 & 88.8\phast  & 69.5\phast  &     \multicolumn{3}{c}{Point-like} & &0.5\\
          & E & 12 36 32.6 &  44 25 09.0 &      & 18.8\phast & \multicolumn{3}{c}{Point-like} &   & \\

1236+444B & W & 12 36 15.6 &  44 27 13.8 &      & 16.4\phast & \multicolumn{3}{c}{Point-like} &   &\\
          & C & 12 36 23.3 &  44 26 11.2 & 38.7\phast &  3.5\phast & \multicolumn{3}{c}{Point-like} &   &  0.5\\

1309+412A & S & 13 09 26.1 &  41 12 09.2 &                & 52.5\phast  & 74.8 &\hspace{-0.3cm}$\times$&\hspace{-0.35cm}52.8 & 29   & \\
          & N & 13 09 28.2 &  41 17 41.8 &   116.9\phast  & 64.4\phast  & 86.3 &\hspace{-0.3cm}$\times$&\hspace{-0.35cm}52.1 & 179  &    1.0\\

1313+387  & S & 13 13 05.8 &  38 45 40.5 &                & 14.6\phast  &  \multicolumn{3}{c}{Point-like}       &      & \\
          & N & 13 13 08.4 &  38 47 22.4 &    45.8\phast  & 31.2\phast  & 30.3 &\hspace{-0.3cm}$\times$&\hspace{-0.35cm}14.2 & 118  &    1.1\\

1318+428C &   & 13 18 59.3 &  42 50 34.5 &                &180.9\phast  & 62.8 &\hspace{-0.3cm}$\times$&\hspace{-0.35cm}35.9 & 110  &\\
1318+428A &   & 13 19 08.9 &  42 51 04.3 &   418.2\phast  &237.3\phast  & 56.1 &\hspace{-0.3cm}$\times$&\hspace{-0.35cm}32.8 & 98   &  1.2\\

1330+380  & W & 13 30 32.0 &  38 01 20.4 &                & 17.6\phast  & 52.0 &\hspace{-0.3cm}$\times$&\hspace{-0.35cm}\ph13.5  &116   & \\
          & E & 13 30 39.8 &  38 00 18.9 &   31.7\phast   & 14.1\phast  &  \multicolumn{3}{c}{Point-like}       &      &   1.0\\

1422+395  & W & 14 22 17.5 &  39 35 26.0 &       & 17.3\phast & 49.3 &\hspace{-0.3cm}$\times$&\hspace{-0.35cm}\ph19.6 & 33 &\\
          & E & 14 22 28.7 &  39 35 15.6 &   33.4\phast& 16.1\phast & 84.8 &\hspace{-0.3cm}$\times$&\hspace{-0.35cm}69.7 & 113 &  0.6\\

1447+402  &   & 14 47 03.1 &  40 13 53.4 & 104.4$^\ast$ &  &   \multicolumn{3}{c}{---} &    &  0.9\\

1450+391B &   & 14 50 09.1 &  39 09 02.7 & 71.7$^\ast$ & &   \multicolumn{3}{c}{---} & & 0.7\\

2303+391A & N & 23 03 43.0 &  39 12 05.8 &   151.1\phast    &     67.0\phast   &    44.5 &\hspace{-0.3cm}$\times$&\hspace{-0.35cm}22.8 &  166 &     0.7\\
          & S & 23 03 44.9 &  39 09 57.8 &           &     84.1\phast   &    64.5 &\hspace{-0.3cm}$\times$&\hspace{-0.35cm}26.1 & 170&\\

2320+416B &   & 23 20 18.6 &  41 41 33.1 &              & 54.1\phast& 26.3 &\hspace{-0.3cm}$\times$&\hspace{-0.35cm}\ph16.0  & 20&\\
2320+417  &   & 23 20 24.2 &  41 43 03.4 &  68.1\phast  & 14.0\phast& \multicolumn{3}{c}{Point-like} &    &  1.0\\

2341+396A &   & 23 41 37.2 &  39 37 18.9 &  20.3\phast  &              &    \multicolumn{3}{c}{Point-like} &  &       1.4\\
2341+396B &   & 23 41 43.5 &  39 35 28.7 &  22.6$^\ast$ &              &   \multicolumn{3}{c}{---} &  &  \\

2351+400B & W & 23 51 22.9 &  40 01 21.8 &             &  98.0\phast   &    58.0 &\hspace{-0.3cm}$\times$&\hspace{-0.35cm}36.1 & 108 &  \\
          & E & 23 51 28.9 &  40 00 57.9 & 157.8\phast &  59.8\phast   &    42.6 &\hspace{-0.3cm}$\times$&\hspace{-0.35cm}40.5 &  27 &     1.6\\

2354+471  &   & 23 54 53.0 &  47 09 13.2 &             & 197.1\phast   &70.2 &\hspace{-0.3cm}$\times$&\hspace{-0.35cm}46.0 &  57 &   \\
2354+471A &   & 23 55 02.4 &  47 10 08.2 & 383.6\phast & 186.5\phast   & 71.7 &\hspace{-0.3cm}$\times$&\hspace{-0.35cm}42.2 &  61 &     0.9\\\hline
\end{tabular}
\end{tiny}
%\end{flushleft}
\end{table*}